\documentclass[12pt]{article} 
\usepackage[paper=letterpaper,margin=.8in]{geometry}

\usepackage{cite}
\usepackage{amsmath}
\usepackage{amssymb}
\usepackage{tensor}
\usepackage{graphicx}
\usepackage{caption}
\usepackage{color}
\usepackage{subfigure}
\usepackage{tikz}
\usepackage[utf8]{inputenc}
\usepackage{braket}
\usepackage{upgreek}
\usepackage[normalem]{ulem}
\usepackage[colorlinks = true, linkcolor = blue, citecolor = red, urlcolor  = blue, anchorcolor = blue]{hyperref}
\usepackage{simpler-wick}
\usepackage{soul}

\newcommand{\calo}{\mathcal{O}}

\newcommand{\disk}{\text{disk}}
\newcommand{\cyl}{\text{cyl}}

\newcommand{\la}{\langle}
\newcommand{\ra}{\rangle}

\newcommand{\wh}{\widehat}

\DeclareMathOperator{\tr}{Tr}

\numberwithin{equation}{section}

\begin{document}

\thispagestyle{empty}

\begin{center}

~\vspace{5mm}

{\LARGE \bf {Matrix models \\ for eigenstate thermalization \\
}}

\vspace{0.5in}

{\bf Daniel Louis Jafferis,$^{1}$ David K. Kolchmeyer,$^{1,2}$ }
\vskip1em
{\bf Baur Mukhametzhanov,$^{3,4}$ and Julian Sonner$^{5}$}

\vspace{0.5in}

$^1$ Department of Physics, Harvard University, Cambridge, MA 02138, USA

\vskip1em

$^2$ Center for Theoretical Physics,

Massachusetts Institute of Technology, Cambridge, MA 02139, USA

\vskip1em

$^3$ Institute for Advanced Study, Princeton, NJ 08540, USA

\vskip1em

$^4$ Department of Physics, Cornell University, Ithaca, NY 14853, USA

\vskip1em

$^5$ Department of Theoretical Physics, University of Geneva, Geneva, Switzerland
    
    \vspace{0.5in}

\end{center}

\vspace{0.5in}

\begin{abstract}

We develop a class of matrix models which implement and formalize the `eigenstate thermalization hypothesis' (ETH) and point out that in general these models must contain non-Gaussian corrections, already in order to correctly capture thermal mean-field theory, or to capture non-trivial OTOCs as well as their higher-order generalizations. We develop the framework of these `ETH matrix models', and put it in the context of recent studies in statistical physics incorporating higher statistical moments into the ETH ansatz. We then use the `ETH matrix model' in order to develop a matrix-integral description of JT gravity coupled to a single scalar field in the bulk. This particular example takes the form of a double-scaled ETH matrix model with non-Gaussian couplings matching disk correlators and the density of states of the gravitational theory. Having defined the model from the disk data, we present evidence that the model correctly captures the JT+matter theory with multiple boundaries, and conjecturally at higher genus.  This is a shorter companion paper to the work \cite{Jafferis:2022wez}, serving both as a guide to the much more extensive material presented there, as well as developing its underpinning in statistical physics.

\end{abstract}

\pagebreak

{
\hypersetup{linkcolor=black}
\tableofcontents
}

%%%%%%%%%%%%%%%%%%%%%%%%%%%%%%%%%
%%%%%%%%%%%%%%%%%%%%%%%%%%%%%%%%%
%%%%%%%%%%%%%%%%%%%%%%%%%%%%%%%%%
%%%%%%%%%%%%%%%%%%%%%%%%%%%%%%%%%
%%%%%%%%%%%%%%%%%%%%%%%%%%%%%%%%%

\section{Introduction}
Theories defined by matrix integrals arise in a variety of physical contexts. Two particularly prominent, but a priori unrelated examples lie at the heart of this paper. On the one hand, such matrix theories arise as the description of ergodic phases of quantum chaotic systems, while on the other hand, they also feature prominently in the study of lower-dimensional quantum gravity where they allow a construction of the non-perturbative sum over (smooth) geometries. Progress in lower-dimensional holography in recent years \cite{Kitaev1,Kitaev2,Kitaev:2017awl,Maldacena:2016upp,Jensen:2016pah,Engelsoy:2016xyb} has led to the striking realization \cite{Cotler:2016fpe} that these two approaches in certain contexts are in fact one and the same. This finds perhaps its clearest expression in the rewriting of the path integral of two-dimensional JT gravity in terms of a matrix integral over boundary Hamiltonians \cite{Saad:2019lba}. The potentially profound connection of two-dimensional gravity to the physics of quantum chaos has been further developed in \cite{Altland:2020ccq,Altland:2022xqx} by demonstrating that JT gravity can indeed be described as a quantum ergodic phase. It is clearly an important goal to develop this connection between quantum chaos and quantum gravity further, in particular by extending it away from pure JT gravity and low dimensions. This work, as well as its longer companion paper, \cite{Jafferis:2022wez}, develop the gravity/chaos correspondence further by devising a matrix-integral description of JT gravity coupled to scalar matter, and furthermore by incorporating a different paradigm of quantum chaos into the story, namely that of the eigenstate thermalization hypothesis (`ETH'). The two-dimensional models we mentioned before may be {\it exactly} rewritten in terms of matrix descriptions of ergodic phases, but one would expect that {\it effective} descriptions of gravity in terms of quantum ergodic matrix theories should also be possible in higher dimensions, at least at or after certain time scales. In conventional quantum-chaotic approaches this time scale would be taken to be the Thouless time, but by adding specific non-linearities or constraints to the matrix integral, one may extend the region of validity to earlier times. The `ETH matrix model' developed in this work and in \cite{Jafferis:2022wez} offers a general framework of the kind of structure that we expect to be relevant also in such cases.

This paper, being a shorter companion to the work \cite{Jafferis:2022wez},  serves two dual purposes. Firstly, we wish to give an accessible and succinct summary of the salient features, results and methods of the longer paper, with the aim to guide the reader through the more extensive technical material appearing there. Secondly, we wish to supply a slightly different perspective on the results, emphasizing general aspects relevant to a statistical description of chaotic quantum systems which unifies random-matrix theory (RMT) and eigenstate thermalization into a common framework. We illustrate the approach by describing how it implements thermal mean-field theory (see Section \ref{eq.TMFT}), the simplest example known to us, as well as to JT gravity which introduces more structure beyond thermal mean field theory (TMFT). 

While our principal object of interest remains the matrix-model description of matter-coupled JT gravity, we believe that the more general structure of the `ETH matrix model' is of interest for more general quantum chaotic systems. At its heart is the simple observation that an ensemble-averaged description even of thermal mean-field theory must contain important non-Gaussian correlations of the form
\begin{equation}\label{eq.nonGausssianIntro}
	\overline{ {\cal O}_{a_1b_1} {\cal O}_{a_2b_2} \cdots {\cal O}_{a_n b_n}} \Bigr|_{\rm ETH} = e^{-(n-1)S(\overline E)}g^{(n)}_{{\cal O}, a_1 b_1 \cdots a_n b_n} \left(E_1,\ldots, E_n  \right) + {\rm disconnected}\,
\end{equation}
which contain terms that are suppressed by further factors of the microcanonical entropy $S(\overline E)$ compared to the usual Gaussian correlations evidenced in the standard `ETH ansatz'.  By $\calo_{a b}$ we mean a matrix element of a simple operator in an energy eigenbasis, and the functions $g^{(n)}_{{\cal O}, a_1 b_1 \cdots a_n b_n} \left(E_1,\ldots, E_n  \right)$ are smoothly varying order-one functions of $n$ energies. We will comment on the index structure of $g^{(n)}_{{\cal O}, a_1 b_1 \cdots a_n b_n} \left(E_1,\ldots, E_n  \right)$ later; for now, we note that in order to contribute to the thermal correlation function at leading order, each $b$ index must be equal to one of the $a$ indices. This explains why $g^{(n)}_{{\cal O}, a_1 b_1 \cdots a_n b_n}$ is a function of only $n$ energies, rather than $2n$ energies.  By ``disconnected'' we are referring to contributions that depend on functions of fewer than $n$ energy arguments such as the smooth functions that appear in the standard Gaussian ETH ansatz. This observation resonates well with recent work on the statistical physics of OTOCs, demonstrating the need for extending ETH to include non-Gaussian contributions (see e.g. \cite{Sonner:2017hxc,Foini:2018sdb,Murthy:2019fgs,Wang:2021mtp}). In this work we formalize the observation above in terms of a two-matrix model of the form
\begin{equation}
	\label{eq:1.2}
{\cal Z}_{\rm ETH} = \int dH d{\cal O} e^{- N {\rm Tr} V(H,{\cal O})}\,,
\end{equation}
where $H$ is a matrix defining the statistics of energy levels, while the ${\cal O}$ matrix generates the correlations of matrix elements of the additional matter field. Below we will be more careful about the definition of the correct integration measure and potentials appearing in the exponent.  Translated to the potential of the ETH matrix model the highly-suppressed non-Gaussianities of \eqref{eq.nonGausssianIntro} appear in fact at ${\cal O}(1)$, which is a reflection of the fact that they must contribute at leading order to higher-order correlation functions in order to produce the necessary connected parts.

The first part of this paper is dedicated to a more careful definition of the ETH matrix model, with general quantum chaotic applications in mind. The second part specializes this structure to the theory of JT gravity with matter, where the additional structure present allows us to go much further in pinning down the full matrix potential $V(H,{\cal O})$, guided by the gravitational path integral as well as arguments based on locality and conformal symmetry. Note that the JT-matter gravitational path integral is not well defined all the way to the UV, unlike the pure JT case. We will point out how this is encoded in the dual ETH matrix model as well.

%%%%%%%%%%%%%%%%%%%%%%%%%%%%%%%%%
%%%%%%%%%%%%%%%%%%%%%%%%%%%%%%%%%
%%%%%%%%%%%%%%%%%%%%%%%%%%%%%%%%%
%%%%%%%%%%%%%%%%%%%%%%%%%%%%%%%%%
%%%%%%%%%%%%%%%%%%%%%%%%%%%%%%%%%

\section{A matrix-model for eigenstate thermalization} \label{MMforETH}

An important signature shown by quantum chaotic systems is the presence of distinctive statistical correlations between their energy levels. There are two interrelated frameworks which are usually invoked to quantitatively describe these correlations, namely Random Matrix Theory (RMT) and the Eigenstate Thermalization Hypothesis (ETH). The purpose of this section is to describe on the one hand a generalization of ETH that is powerful enough to incorporate more fine-grained correlations than the usual `Gaussian ansatz', and at the same time to introduce a random-matrix description of such a generalized ETH ensemble. The latter will set the general context of our matrix-model description of JT gravity coupled to simple scalar matter, described in Section \ref{sec.JTmatterMatrixModel}, as well as in the longer companion paper \cite{Jafferis:2022wez}. 
\subsection{Degaussing ETH}\label{sec.deGaussETH}
The level correlations implied by ETH are usually stated in the form of an ansatz for the matrix elements of a simple operator in the energy basis \cite{Deutsch,Srednicki1},
\begin{equation}\label{eq.GaussianETH}
\langle E_a | {\cal O} | E_b \rangle = \overline{\langle E_a | {\cal O} | E_a \rangle} \delta_{ab} + e^{-S(\overline E)/2} f_{\cal O}\left(E_a, E_b\right) R_{ab}\,,
\end{equation}
where $\overline{\langle E_a | {\cal O} | E_a \rangle}$ is a smooth function of the average energy $\overline{E} = \frac{1}{2}\left(E_a + E_b \right)$, $S$ is the microcanonical entropy evaluated at $\overline{E}$ and $f_{\cal O}\left(E_a, E_b\right)$ is a smooth function of both energies. Finally $R_{ab}$ is a random matrix, traditionally taken to be sampled from a {\it Gaussian} distribution with zero mean and unit variance. One therefore postulates the existence of an ensemble -- we will refer to this as the `ETH ensemble' -- from which matrix elements of simple operators in the energy basis are sampled. The usual ETH ansatz above is equivalent to specifying the first two non-trivial moments of this ensemble,
 namely compatible with a purely Gaussian probability distribution. 

That this ensemble cannot in fact obey purely Gaussian statistics can be seen by noting that this would imply that all higher-order thermal correlation functions factorize, which is in contradiction with, for example, the existence of a non-vanishing Lyapunov exponent diagnosed from OTOCs in eigenstates \cite{Sonner:2017hxc,Foini:2018sdb,Murthy:2019fgs,Anous:2019yku,Nayak:2019evx}. Another argument forcing us to consider non-Gaussian statistical ensembles comes from considering the main application we have in mind in this work, namely CFTs with holographic duals, and more specifically JT gravity coupled to matter. Suppose we use bulk gravity Witten-diagram techniques to compute the manifestly crossing-invariant four-point function
\begin{equation}
 \braket{\calo(\tau_1) \calo(\tau_2) \calo(\tau_3) \calo(\tau_4)} = 
 (\tau_{12}\tau_{34})^{-2 \Delta} 
 +  (\tau_{14}\tau_{23})^{-2 \Delta}
+ (\tau_{13}\tau_{24})^{-2 \Delta}
\label{eq:rightGFFcorrelator}
\end{equation}
of the boundary 1D conformal quantum system. We may also consider its generalization to finite temperature
\begin{align}
\braket{ \tr e^{-\beta H }\calo(\tau_1) \calo(\tau_2) \calo(\tau_3) \calo(\tau_4)} = 
 &\left(
{\beta \over \pi} \sin( {\pi \over \beta} \tau_{12}) ~  {\beta \over \pi} \sin( {\pi \over \beta} \tau_{34})
 \right)^{-2 \Delta} \\
 + &
  \left(
{\beta \over \pi} \sin( {\pi \over \beta} \tau_{14}) ~  {\beta \over \pi} \sin( {\pi \over \beta} \tau_{23})
 \right)^{-2 \Delta} \\
+&
 \left(
{\beta \over \pi} \sin( {\pi \over \beta} \tau_{13}) ~  {\beta \over \pi} \sin( {\pi \over \beta} \tau_{24})
 \right)^{-2 \Delta}  \ .
\label{eq:rightGFFcorrelatorFT}
\end{align}
By crossing invariance we refer to the invariance of the four-point function under permutations of the external operators. Using the bulk perspective, the expression above results straightforwardly from a computation using gravity Witten diagrams, so long as we work in the semi-classical regime. From the boundary perspective, this expression implies that the OPE between ${\cal O}(x) {\cal O}(y)$ only receives contributions from double trace primaries $\left[ {\cal O O} \right]_n$ of dimension $2\Delta + n$ for $n \in 2 \mathbb{Z}_{\ge 0}$ in addition to the identity operator. The latter perspective may not seem to be immediately relevant to the discussion, but we shall come to this statement shortly, and put it to good use in the context of  thermal mean field theory and indeed JT gravity, where it will be of great help to constrain the relevant ETH ensemble.

To set the scene, we now ask what the answer would be if we evaluated the correlation function above in the Gaussian ETH ensemble, i.e using the ansatz \eqref{eq.GaussianETH} to define quadratic Wick contractions of the operator\footnote{For simplicity, let's focus on the case where the operator has vanishing one point function, so that $\overline{\langle E_a | {\cal O} | E_a \rangle}=0$. The result can obviously be generalized to incorporate a non-trivial one-point function as well. } ${\cal O}$. We thus compute
\begin{eqnarray}\label{eq.GaussianETH4pt}
\overline{\text{Tr } \,\, \prod_{j=1}^4 e^{-\beta_j H}{\cal O}}\Biggr|_{\rm Gauss}  &=& \sum_{a_1 \ldots a_4} e^{-\sum_j \beta_j E_{a_j}} F(E_{a_1}, E_{a_2})^{-1} F(E_{a_3}, E_{a_4})^{-1} \left(  \delta_{a_1 a_3}  + \delta_{a_2 a_4}\right)\\
&&+ \sum_a e^{-(\beta_1 + \cdots + \beta_4) E_a} F(E_a,E_a)^{-2}\,.
\label{eq.GaussianETH4pt2}
\end{eqnarray}
Note that, as indicated, we evaluated the above expectation value with respect to the purely Gaussian ETH ensemble, using the Wick contraction
\begin{equation}
\wick{ \c1 { \cal O}_{ab} \c1 {\cal O}_{cd} }= e^{-S(\overline{E})} \left|  f_{\cal O}(E_a, E_b)\right|^2 \delta_{ad} \delta_{bc}\, := F(E_a, E_b)^{-1} \delta_{ad} \delta_{bc}
\end{equation}
following from the Gaussian ETH ansatz above, and where in the last equality we have introduced a more convenient notation that will be helpful later on. The average energy in the exponent reads $\overline{E} = \frac{1}{2} \left(E_a + E_b  \right)$. We now notice that the third term \eqref{eq.GaussianETH4pt2} is non-planar\footnote{Non-planar here in the sense of matrix-model perturbation theory: we draw all index contractions using 't Hooft double-line notation, resulting in a ribbon graph whose topology is called `planar' whenever it can be drawn on a sphere, and non-planar otherwise.}, and as such is sub-leading at large $N = {\rm dim} H \gg 1$, or, equivalently, in the $e^S$ counting. 

Considering only the first two terms which contribute at leading order, however, gives an answer that is not crossing invariant, equivalent to only keeping the first two terms in \eqref{eq:rightGFFcorrelator} above. We conclude that the term making the overall result invariant must come from a non-Gaussian connected contribution to the ETH ensemble, which nevertheless contributes at leading order owing to additional Hilbert space sums, i.e. a contribution of the form
\begin{equation}
\overline{{\cal O}_{a_1b_1} {\cal O}_{a_2b_2} {\cal O}_{a_3 b_3} {\cal O}_{a_4b_4}} \supset e^{-3 S(\overline E)}g^{(4)}_{{\cal O}, a_1 b_1 \cdots a_4 b_4} \left(E_1,E_2,E_3,E_4  \right) \,,
\end{equation}
where $\overline{E}$ is now the average of all four energies, and $g_{\cal O}$ is a smooth function of the individual energies. The entropic suppression factor is dictated by the need to precisely balance the four sums over energy eigenstates in the four-point function \eqref{eq.GaussianETH4pt} to result in a leading order contribution. For a general $n-$point function this generalizes easily to
\begin{equation}\label{eq.GeneralNonGaussianity}
\overline{ {\cal O}_{a_1b_1} {\cal O}_{a_2b_2} \cdots {\cal O}_{a_n b_n}} = e^{-(n-1)S(\overline E)}g^{(n)}_{{\cal O}, a_1 b_1 \cdots a_n b_n} \left(E_1,\ldots, E_n  \right) + {\rm disconnected}\,.
\end{equation}
In typical expressions of interest the above averaged matrix elements will be integrated against a suitable set of densities of states, which we will incorporate below into our construction. The notation of the overline $\overline{\cdots}$ indicates purely the expectation value with respect to the ensemble determining matrix elements, while $\langle \cdots \rangle$ will indicate an expectation value that includes the energy integrals, a structure we refer to as the `ETH matrix model' for reasons that will become clear presently.

Returning to our discussion of the crossing-invariant four-point function above, we now see that the third planar, i.e. leading order in $e^{S}$, contribution to \eqref{eq:rightGFFcorrelator} comes from the quartic non-Gaussianity $g^{(4)}_{\cal O}(E_1,E_2,E_3,E_4)$, which shows that indeed the ETH ensemble has to contain non-trivial higher statistical moments even to match thermal mean-field theory at the level of the four-point function. At the level of the usual ETH ansatz these non-Gaussianities are extremely subtle, that is highly suppressed in entropy (cf. \eqref{eq.GeneralNonGaussianity}), but their effect on thermal correlation functions is significant and necessary even to ensure basic properties like crossing symmetry.

In the formula \eqref{eq.GaussianETH}, traditionally referred to as the ETH ansatz, the function $f_{\cal O} (E_1,E_2) = f_{\cal O}(\overline{E},\omega)$ is left unspecified, reminding us that it depends on a given physical system. We take the same point of view regarding the higher moments $g^{(n)}_{\cal O}$ in general. In Section \ref{sec.JTmatterMatrixModel}, when we construct the specific ETH ensembles describing TMFT, as well as matter-coupled JT gravity, we are able to fully determine (at least in principle) these non-Gaussianities.

We now comment on the index structure of $g^{(n)}_{{\cal O}, a_1 b_1 \cdots a_n b_n} \left(E_1,\ldots, E_n  \right)$ and specify what we precisely mean by the `generalized ETH ansatz.' Assuming no degeneracy in the spectrum of the Hamiltonian, the energy eigenbasis is only defined up to an energy dependent rotation, $\ket{E_a} \rightarrow e^{i \theta_a} \ket{E_a}$. Because \eqref{eq.GeneralNonGaussianity} should hold for any choice of the phases $\theta_a$, the terms on the right side should contain Kronecker delta symbols that each involve one $a$ and one $b$ index. We also want our ansatz to be invariant under unitary rotations that act within microcanonical Hilbert spaces. That is, we want
\begin{equation}
\sum_{\tilde{a}_1 \tilde{b}_1 \cdots \tilde{a}_n \tilde{b}_n} U\indices{_{a_1} _{\tilde{a}_1}}U^\dagger_{\tilde{b}_1 b_1} \overline{{\cal O}_{\tilde{a}_1\tilde{b}_1}  \ldots  {\cal O}_{\tilde{a}_n\tilde{b}_n}} U_{a_n \tilde{a}_n} U^\dagger_{\tilde{b}_n b_n}= \overline{{\cal O}_{a_1b_1}  \ldots {\cal O}_{a_nb_n}}  \,,
\end{equation}
where $U$ is a block-diagonal unitary where each block has size $e^{S(E)}$. To achieve this, we will define our generalized ansatz such that there are exactly $n$ Kronecker symbols per term. In particular, in our ansatz we exclude symbols involving two $a$ or two $b$ indices. The general index structure is then
\begin{equation}
	\label{eq:indices}
	\delta_{a_1 b_{\sigma(1)}} 
	\delta_{a_2 b_{\sigma(2)}}
	\cdots
	\delta_{a_n b_{\sigma(n)}},  
\end{equation}
where $\sigma \in S_n$ is a permutation of $n$ elements. If $\sigma$ has more than one cycle, then we say that \eqref{eq:indices} factors into several terms that individually take the form of \eqref{eq:indices} for lower values of $n$. Appealing to minimality, we define our generalized ETH ansatz such that a factor with $n$ symbols is uniquely associated to a specific function of $n$ energies. Hence, the generalized ETH ansatz for a single simple operator is completely characterized by a single smooth function of $n$ energies for each $n$. In particular, we have \begin{equation}
g^{(n)}_{{\cal O}, a_1 b_1 \cdots a_n b_n} \left(E_1,\ldots, E_n  \right) = \sum_{\sigma \in S_n} g^{(n)}_{{\cal O}} \left(E_{\sigma(1)},\ldots, E_{\sigma(n)}  \right) \delta_{b_{\sigma(1)} a_{\sigma(2)}} \delta_{b_{\sigma(2)} a_{\sigma(3)}} \cdots \delta_{b_{\sigma(n)} a_{\sigma(1)}} \, ,
\label{eq:refinedansatz}
\end{equation}
where $\sigma$ is a permutation of $\{1,2,\cdots,n\}$, and $g^{(n)}_{{\cal O}}(E_1,E_2,\cdots,E_n)$ is, w.l.o.g., a cyclically invariant function of $n$ energies. Our `generalized ETH ansatz' is the minimal modification of the standard Gaussian ETH ansatz that is needed for compatibility with thermal mean field theory, and is completely characterized by the functions $g^{(n)}_{{\cal O}} \left(E_1,\ldots, E_n  \right)$ for all $n \in \mathbb{N}$. These functions are theory-dependent, and in JT gravity it is easy to deduce them from the expressions for the disk correlators.\footnote{In JT gravity minimally coupled to a free scalar field, $n$ will only take on even values, owing to the $\calo \rightarrow - \calo$ symmetry.} This structure is naturally produced by a two-matrix model with a single-trace matrix potential, which is why the potential in \eqref{eq:1.2} is written with a single trace.

\subsection{Matrix-model description}
One of the main insights about chaotic quantum systems is the (conjectured and empirically robustly observed) fact that their energy eigenvalues follow correlation laws associated to those of random matrix ensembles \cite{bohigas1984characterization}. One defines the joint probability distribution of the eigenvalues of a random matrix \cite{mehta2004random}
\begin{equation}\label{eq.DysonBetaMeasure}
d\mu[H] = \mu_0^{-1}e^{-\sum_a V(E_a)} \prod_{a<b} \left|E_a - E_b\right|^\beta\,,
\end{equation}
where $\beta \in \left\{1,2,4 \right\}$ depends on the symmetry class\footnote{We show formulae for the three so-called classical ensembles for $\beta =1$ (GOE), $\beta = 2$ (GUE) and $\beta =4$ (GSE).} and $\mu_0$ is a normalization factor, whose value we do not need here. In our principal application to JT with matter, we will be interested in the case $\beta=2$, namely when the matrix model is in the unitary class, but the remaining cases  are relevant for more general ETH ensembles as well as other bulk theories, see e.g. \cite{Stanford:2019vob}. In a number of standard applications, the potential $V(E_a)$ is taken to be quadratic, but we will be interested in higher-order non-Gaussian generalizations of this, for example the so-called SSS potential, describing the eigenvalue density of JT gravity in terms of a matrix integral \cite{Saad:2019lba}. Integrating this probability density over all but $n$ eigenvalues gives the $n-$level density, or $n-$level correlation function, \cite{mehta2004random},
\begin{equation}
\rho^{(n)} (E_1,\ldots E_n) = \rho(E_1)\cdots \rho(E_n) + {\rm connected}\,.
\end{equation}
A special case of this is the spectral density, obtained by marginalizing the distribution \eqref{eq.DysonBetaMeasure} over all but a single eigenvalue. We often use the notation
\begin{equation}
\rho(E) = e^{S(E)} = e^{S_0} \rho_0 (E)\,,
\end{equation}
meaning that we alternately write the full spectral density as the exponent of the microcanonical entropy at a given energy $E$, or as a factor $e^{S_0}$ times a smooth order-one function $\rho_0(E)$, as is often employed in the SYK / JT context, where $S_0$ has the meaning of the ground-state entropy \cite{Sachdev:2015efa,Maldacena:2016hyu,Jensen:2016pah, Maldacena:2016upp}. More generally, we can think of $e^{S_0}$ as a bookkeeping parameter of entropic factors.
 Putting together the RMT description of the energy levels of chaotic systems with the generalized ETH of section \ref{sec.deGaussETH} above, makes it very natural to describe the overall structure in terms of a two matrix model where the energy-level statistics are generated by the random Hamiltonian $H$, while the operator matrix element statistics (the generalised ETH) are generated by a second random matrix, which we denote with the same symbol ${\cal O}$, by a slight abuse of notation.  The matrix integral
\begin{equation}
{\cal Z}_{\rm ETH} = \int d\mu[H] d\mu[{\cal O}] e^{- {\cal V}\left[H,{\cal O}  \right]}
\end{equation}
is then seen as the joint probability distribution of energy levels and matrix elements of this ensemble with $d\mu[H] d\mu[{\cal O}]$ the yet to be specified appropriate measures for the two random matrices, but which we can already anticipate will generally neither be quadratic in $H$ nor in ${\cal O}$. Viewed in the energy eigenbasis this is exactly the `ETH ensemble' we have previously invoked. In this section we will describe the general features of such an ensemble, before constructing a specific instance in Section \ref{sec.JTmatterMatrixModel}, capable of describing matter-coupled JT gravity.

In view of the ETH ansatz and its beyond-Gaussian generalization it is most natural to define the ETH matrix model in the energy eigenbasis of the Hamiltonian with a single-trace matrix potential
\begin{equation}\label{eq.ETHTwoMatrixModel}
{\cal Z}_{\rm ETH} = \int d\mu[H] d\mu[{\cal O}] \exp\left( e^{S_0}\sum_{n \ge 2} \sum_{a_1 \cdots a_n}G^{(n)}(E_{a_1},\cdots,E_{a_n})  {\cal O}_{a_1 a_2} {\cal O}_{a_2 a_3} \cdots  {\cal O}_{a_n a_1}\right)\,,
\end{equation}
with the measure for the energy eigenvalues defined to be \eqref{eq.DysonBetaMeasure} above, while 
\begin{equation}
d\mu [{\cal O}] :=  \prod_i d{\cal O}_{ii}\prod_{i<j} d {\rm Re}{\cal O}_{ij} d {\rm Im}{\cal O}_{ij}\,.
\end{equation}
The $G^{(n)}$ functions encode the $g_\calo^{(n)}$ functions introduced in \eqref{eq:refinedansatz}, and their precise relationship can only be determined by actually performing the matrix integral above, which is difficult in general. The reader may want to convince themselves that this matrix model results in the necessary $e^{S_0}$ scaling shown in \eqref{eq.GeneralNonGaussianity} if the matrix-model coupling functions $G^{(n)}$  are of ${\cal O}(1)$ in $e^{S_0}$. This structure is thus a matrix-integral representation of the generalized ETH above, capable of producing at the same time the statistical distribution of energy eigenvalues and corresponding spectral integrals, as well as the statistics of the operator matrix elements ${\cal O}_{ab}$. The coupling functions $G^{(n)}(E_1,\ldots E_n)$ are smooth functions of the energy arguments, and take on specific functional forms only once a particular theory has been specified. As an example, we will indicate below (again much more detail can be found in \cite{Jafferis:2022wez}) how these functions can in principle be determined systematically for the JT+scalar matrix model from thermal correlation functions.

Note that $d\mu[H]$, as defined in \eqref{eq.DysonBetaMeasure} above, contains both the Vandermonde factor, $\Delta^\beta(E)$, as well as the exponential of the potential $V(E_a)$. The joint measure can be seen to be the correct one by following the usual Fadeev-Popov procedure \cite{Marino:2004eq}. One starts in a general basis where neither $H$ nor ${\cal O}$ are diagonal and then transforms to the eigenbasis of the $H-$matrix, sending $H\rightarrow U H U^\dagger$ and ${\cal O} \rightarrow U {\cal O} U^\dagger$. In general $H$ and ${\cal O}$ are not simultaneously diagonalisable, so that we may reduce to integrating only over the eigenvalues of $H$, but must integrate over the full Haar measure for Hermitian matrices for the second matrix ${\cal O}$.

As mentioned before, the $G^{(n)}(E_1,\ldots , E_n)$ are continuous functions of the energies  and they determine the matrix model potential, which is organized as an expansion in ${\cal O}^n$. We will often write the $n-$th order term in this expansion as $W^{(n)} \left[E_{a_1},\ldots E_{a_n}; {\cal O}  \right] $. One chooses $V(H)$ in such a way as to match the leading density of states to the chaotic system of interest\footnote{Note that a matrix model will result in a continuous density of states. We may view this as a coarse-grained density of state of a general quantum chaotic system, or alternatively take this density as part of the definition of the model, as will be the case in the application to minimally coupled JT gravity.}. We can now deduce the statistical distribution of energy eigenvalues, as well as of operator matrix elements by introducing sources for the energy, as well as for moments of the operator ${\cal O}$ in the energy eigenbasis in order to generate the various moments of the ETH ensemble -- these will of course be nothing but our $g^{(n)}_{{\cal O}, a_1 b_1\ldots a_n b_n}(E_1,\ldots E_n)$ above. Let us illustrate this for the quadratic and quartic moments before stating the general answer. From the model \eqref{eq.ETHTwoMatrixModel} one computes the two-point function of two operators at arbitrary Euclidean separation
\begin{eqnarray}\label{eq.ETHMatrixModelTwoPoint}
\left\langle{\rm Tr} \left(e^{-\beta_1 H}{\cal O} e^{-\beta_2 H}{\cal O} \right) \right\rangle &=& \int d\mu[H] d\mu[{\cal O}]  ~ {\rm tr} \left(e^{-\beta_1 H}{\cal O} e^{-\beta_2 H}{\cal O} \right)  e^{ e^{S_0}\sum_{n} W^{(n)} \left[E_{a_1},\ldots E_{a_n}; {\cal O}  \right] }\nonumber\\
&=& \int  dE_a dE_b ~ \rho(E_a) \rho(E_b) e^{-\beta_1 E_a - \beta_2 E_b} \overline{{\cal O}_{ab} {\cal O}_{ba}} + \cdots
\end{eqnarray}
where in the second line we used the overbar notation to indicate the expectation value evaluated in the ${\cal O}$ matrix potential, written in the $H$ eigenbasis. We have consequently performed all energy integrals except for the pair $E_a, E_b$ appearing explicitly in $ \overline{{\cal O}_{ab} {\cal O}_{ba}}$. This has produced a factor of the pair correlation function $\rho^{(2)} \left( E_a, E_b \right)$, which at leading order equals the product $\rho(E_a) \rho(E_b)$. The corrections are subleading in the `genus-expansion', that is in powers of $e^{S_0}$. We may denote this equation graphically as
\begin{align}
\raisebox{-.3in}{\includegraphics[scale=.4]{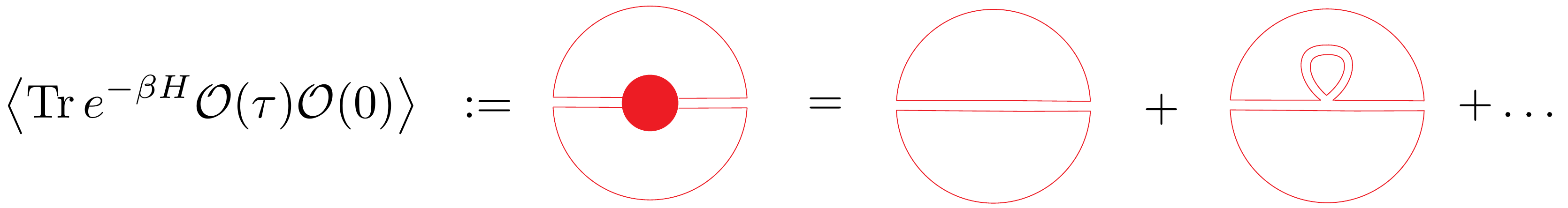}}\,
\label{eq:interactingtwopointthooft}
\end{align}
where we have not drawn any of the in-filling $H$ double lines, which are automatically accounted for by the integrals over the densities of states, as explained above. We may similarly compute the four-point function, which for simplicity we only represent graphically,
\begin{equation}
\includegraphics[scale=0.38]{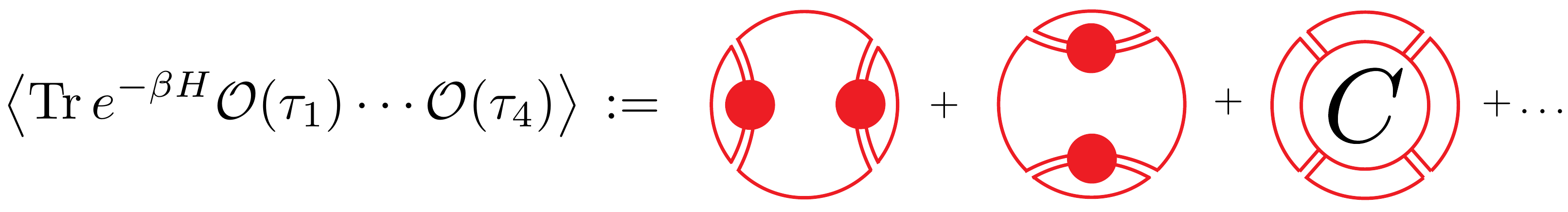}\,
    \label{eq:9.7}
\end{equation}
where the third diagram marked with a ``C" implements the contribution of the interaction terms in the matrix model. Of course, analogous computations apply for higher-order correlations. In any given application one must choose the couplings of the matrix model, as well as the density of states, so as to match the physical correlators of interest. As we will show in section 3, we can do this for the JT-matter matrix model, where we fix the full ETH matrix model by matching to the planar correlation functions. We also provide some evidence for our conjecture that the resulting model reproduces the higher topology correlation functions as well.

\subsubsection*{ETH and topology}
\begin{figure}
	\centering
	
$$	\raisebox{-1in}{\includegraphics[scale = .2]{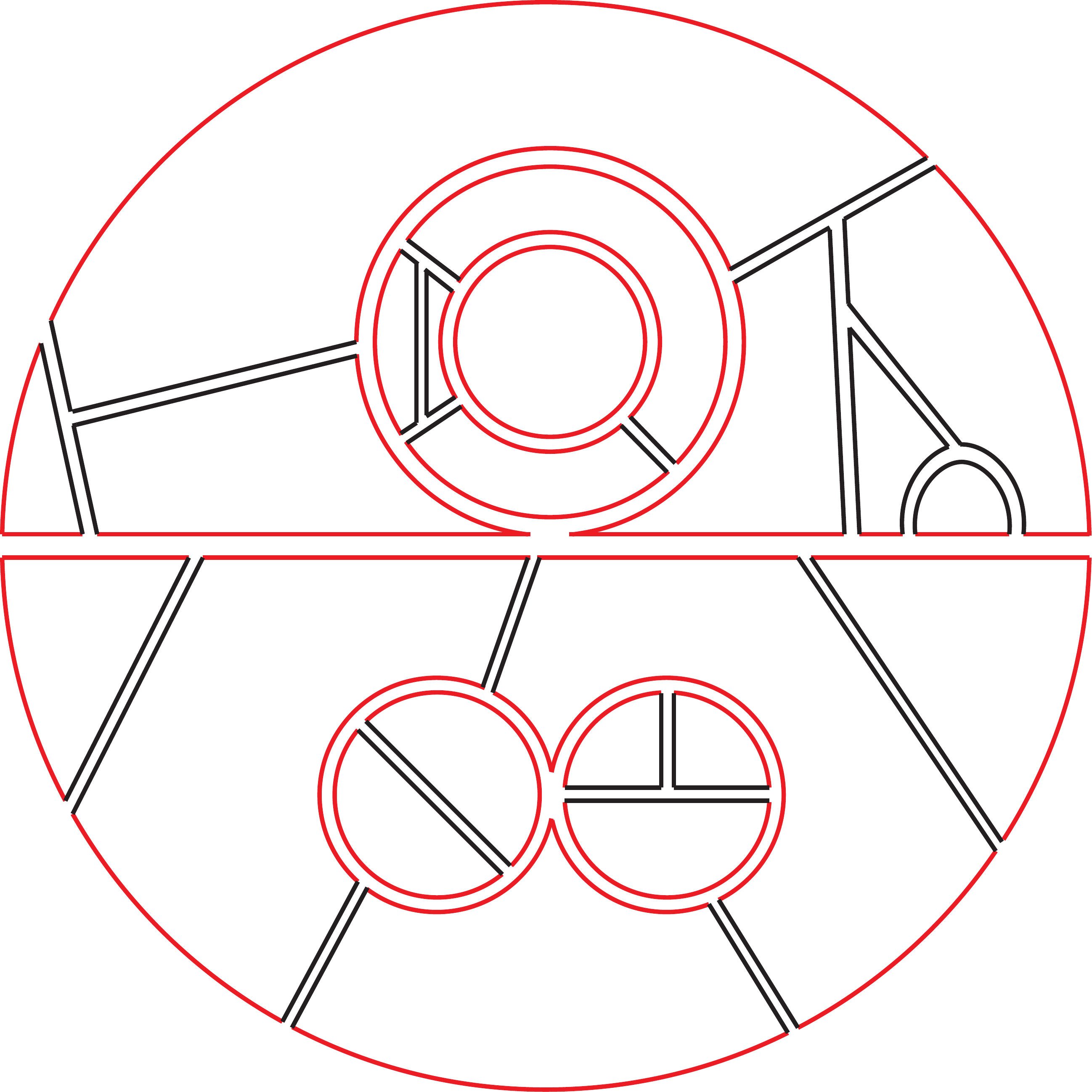} }\sim {\cal O}\left( e^{S_0} \right)\hskip2cm 	\raisebox{-1in}{\includegraphics[scale = .2]{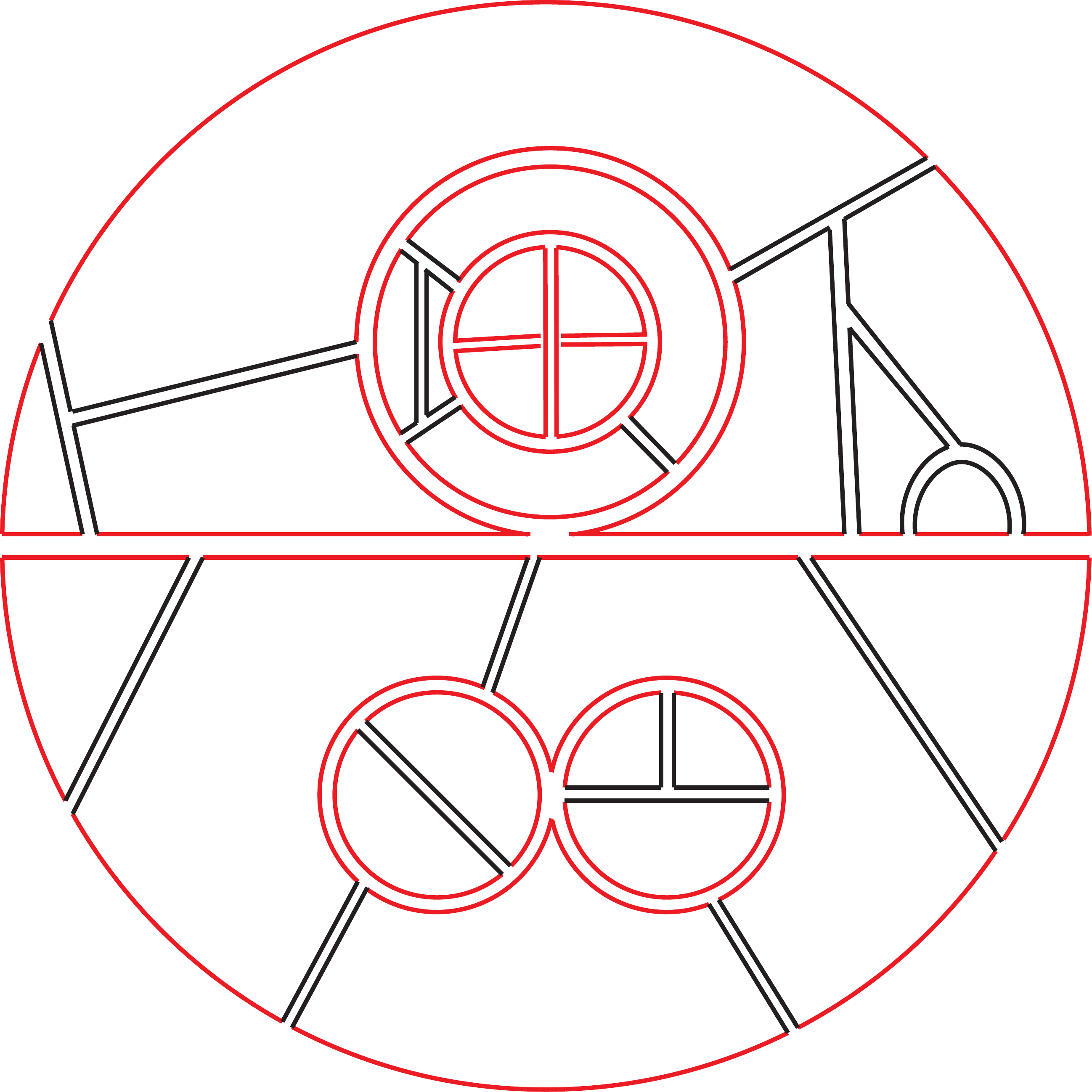}} \sim{\cal O} (e^{-S_0}) $$
	\caption{Topology of the ETH matrix model. The left panel shows a planar diagram contributing to the two-point function of operators \eqref{eq.ETHMatrixModelTwoPoint} which, un-normalized, gives an answer of order $e^{S_0}$. Any non-planar diagram, exemplified by the diagram on the right, contributes at lower-order in the $e^{S_0}$ expansion, the example above contributing at ${\cal O}\left( e^{-S_0} \right)$. The order of an arbitrary diagram is determined by its Euler characteristic $e^{\chi S_0 }$ when viewing its double-line representation as a triangulation of a 2D Riemann surface. This well-known feature of matrix models is the reason why non-Gaussian contributions in the matrix model action are needed for the ETH matrix model in order to reproduce basic facts even about thermal mean-field theory.} 
	\label{fig:ETHTopology}
\end{figure}
From the perspective of generalized eigenstate thermalization, that is non-Gaussian ETH, various entropic factors $e^{S}$ (or equivalently the bookkeeping parameter $e^{S_0}$) are mere kinematical necessities imposed on us by the requirement of producing the correct (e.g. crossing-invariant) four- (and higher-) point functions {\it at leading order}, i.e. they arise from the need to compensate the multiple sums over Hilbert space occurring in expressions such as  \eqref{eq.GaussianETH4pt}. On the other hand, the matrix-model description developed here, gives an alternative interpretation in terms of topology. One may arrange the various matrix-model diagrams contributing to a given correlation function in a 't Hooft expansion and organize the powers of $e^{S_0}$ in terms of the topology shared by all diagrams at a given order (see Figure \ref{fig:ETHTopology}). It is a classic result by t'Hooft \cite{tHooft:1973alw} that the power of $e^{S_0}$ in a diagram is given by
$ e^{\chi S_0}$, whee $\chi = 2-2g -b$ is the Euler characteristic of the surface on which the diagram can be drawn without self-intersections, $g$ is the genus (number of handles) of the surface, and $b$ is the number of boundaries. This is deduced from the ribbon-graph representation by the familiar formula $2-2g= V-E+F$ in terms of the number of (V)ertices, (E)dges and (F)aces. 
 This is of course a well known feature of matrix models (and two-dimensional gravity), but its relation to the non-Gaussianities occurring in extensions of the eigenstate thermalization hypothesis \cite{Foini:2018sdb} is useful and, to the best of our knowledge, new. Connections between topological expansion and quantum chaos have  been described recently \cite{Saad:2019lba,Saad:2019pqd,Pollack:2020gfa,Altland:2020ccq,Cotler:2020ugk,Cotler:2020lxj,Altland:2021rqn,Altland:2022xqx}, largely motivated by the occurrence of so-called `wormhole' solutions contributing to the Euclidean quantum gravity path integral. It is pleasing to be able to incorporate this topological structure into the quantum chaos / ETH discussion in a natural way. As mentioned above, we define our generalized ETH ansatz using a single-trace matrix potential. Single-trace matrix models admit an interpretation where the physical degrees of freedom that live on the two-dimensional surfaces interact locally. The local nature of JT gravity minimally coupled to a free scalar field further motivates us to consider only single-trace matrix models.

\label{sec:ETHmatrixmodel}

\section{Example 1: Thermal Mean Field Theory}\label{eq.TMFT}

As a simple example, we would like to give a matrix model description of a Thermal Mean Field Theory (TMFT), in which correlators factorize into two-point functions. An example of this is the Generalized Free Field arising in the semi-classical limit $G_N \to 0$ of JT gravity coupled to a free massive scalar field, that we will consider in the next section. However, the discussion in this section is more general and applies to TMFT with an arbitrary two-point correlator and an arbitrary density of states.

\bigskip

Thermal Mean Field Theory is determined by its two-point correlator
\begin{align}
{\cal G}_\beta(\tau) &= {1\over Z(\beta)} \tr e^{-\beta H} {\cal O} (\tau) {\cal O} (0)   \\
&= \int_{-\infty}^\infty d\omega ~ e^{\tau \omega} ~ e^{-{\beta\over 2}\omega} \wh {\cal G}_\beta (\omega) \ .
\label{Ghatdef}
\end{align}
In the second line we defined the Laplace transform $\wh {\cal G}_\beta(\omega)$. The integral over $\omega$ is presumed to converge for $0 < \tau < \beta$, which implies that $\wh {\cal G}_\beta(\omega)$ must decay sufficiently rapidly at large $\omega$: $\wh G_\beta(\omega) \lesssim \# e^{-{\beta \over 2}|\omega|}, |\omega| \to \infty$. In \eqref{Ghatdef} we introduced an explicit factor $e^{-{\beta \over 2} \omega}$, such that the KMS condition takes a simple form
\begin{align}
\text{KMS}: \qquad {\cal G}_\beta(\tau) = {\cal G}_\beta(\beta - \tau) \quad \Leftrightarrow  \quad 
\wh {\cal G}_\beta(\omega) = \wh {\cal G}_\beta(-\omega) \ .
\end{align}

\bigskip

The four-point correlator in TMFT factorizes into two-point functions 
\begin{align}
{\cal G}^{(4)}_\beta(\tau_1, \dots, \tau_4) &= 
{1\over Z(\beta)} \tr e^{-\beta H} {\cal O} (\tau_1) \dots {\cal O} (\tau_4) \\
&={\cal G}_\beta(\tau_{12}){\cal G}_\beta(\tau_{34})
+ {\cal G}_\beta(\tau_{14}){\cal G}_\beta(\tau_{23})
+{\cal G}_\beta(\tau_{13}){\cal G}_\beta(\tau_{24}) \ ,
\label{4ptTMFTfact}
\end{align}
where $\tau_{ij} = \tau_i - \tau_j$ and we assume $\beta > \tau_1 > \tau_2 > \tau_3 > \tau_4>0$. Similarly, higher-point correlators are given by a sum of all Wick contractions.

In order to discuss the matrix model for TMFT, it is convenient to first write the correlators in the energy basis
\begin{align}\label{2ptTMFT}
{\cal G}_\beta(\tau) &= {1\over Z(\beta)} \int dE_1 dE_2 ~ \rho(E_1) \rho(E_2) ~ e^{-(\beta-\tau) E_1 - \tau E_2} ~ \la |{\cal O}_{E_1E_2}|^2 \ra \ , \\
{\cal G}^{(4)}_\beta(\tau_1, \dots, \tau_4) &= {1\over Z(\beta)}
\int \prod_{j=1}^4 (dE_j ~ \rho(E_j) e^{-\beta_j E_j}) ~ 
\la {\cal O}_{E_1E_2} {\cal O}_{E_2E_3} {\cal O}_{E_3E_4} {\cal O}_{E_4E_1} \ra \ ,
\label{4ptTMFT}
\end{align}
where 
\begin{align}
    \beta_1 = \beta - \tau_{14}, \qquad  
    \beta_2 = \tau_{12}, \qquad 
    \beta_3 = \tau_{23} , \qquad 
    \beta_4 = \tau_{34} \ .
\end{align}
The integrals over energies are done with the density of states $\rho(E)$. This of course depends on the particular system and should be given in addition to the two-point function to define TMFT. In particular, the partition function is given by the usual expression
\begin{align}\label{Pfunction}
Z(\beta) = \int dE ~ \rho(E) e^{-\beta E} \ .
\end{align}

We will work in the thermodynamic limit. This could be either a large volume limit or, more generally, large number of degrees of freedom, while keeping the temperature fixed. In holographic systems, such as JT gravity, this corresponds to the semi-classical gravity limit $G_N \to 0$. The average energy of the system in the thermodynamic limit is large, typically proportional to the volume or the number of degrees of freedom. In practice, this means that energy integrals, such as \eqref{2ptTMFT}, \eqref{4ptTMFT}, and \eqref{Pfunction}, are dominated by large energies $E_j$, while the differences are much smaller, $E_j \gg |E_i-E_k|$.

For example, the partition function is dominated by the saddle energy determined by the usual thermodynamic relation
\begin{align}
S'(E) = \beta \quad \Rightarrow \quad  E = E(\beta) \ ,
\end{align}
where $\rho(E) = e^{S(E)}$. Also including fluctuations around the saddle, we have
\begin{align}
Z(\beta) &\approx e^{S(E(\beta)) - \beta E(\beta)}
\int d\omega ~ e^{S''(E) {\omega^2 \over 2}}\\
& =e^{S(E(\beta)) - \beta E(\beta)}
\int d\omega ~ e^{- {\beta^2 \over C} {\omega^2 \over 2}} \\
& = {\sqrt{2\pi C} \over \beta} ~ e^{S(E(\beta)) - \beta E(\beta)} \ ,
\end{align}
where we used the standard thermodynamic relation $S''(E) = {dT \over dE} {d \over dT} S'(E) = - {1\over C T^2}$ and $C = {dE \over dT}$ is the heat capacity. In the thermodynamic limit the heat capacity $C$ is typically proportional to the volume or number of dof and therefore $C \to \infty$. Similarly, further non-gaussian fluctuations are suppressed.

\subsection{Two-point function}

Now we consider \eqref{2ptTMFT}, \eqref{4ptTMFT} and higher-point correlators in TMFT. Given the two-point function \eqref{Ghatdef}, we would like to determine the correlators of $\cal O$ in the energy basis. Then we will discuss the matrix model that computes these correlators.

We start with the two-point function. It turns out that
\begin{align}\label{2ptETMFT}
\la |{\cal O}_{E_1 E_2}|^2 \ra = 
{1\over \rho(E)} ~ \widehat{\cal G}_{\beta(E)}(\omega) \ ,
\end{align}
where the inverse temperature is determined by the thermodynamic relation $\beta(E) = S'(E)$ and
\begin{align}
E= {E_1 + E_2 \over 2}  \ , \qquad \omega = E_1 - E_2 \ .
\end{align}
This is valid in the thermodynamic limit. Let's check that it holds. We insert \eqref{2ptETMFT} into \eqref{2ptTMFT}. We switch to coordinates $E,\omega$ and use that
\begin{align}
\rho(E_1) \rho(E_2) &= e^{S(E+{\omega \over 2}) + S(E-{\omega \over 2})} \\
& = e^{2S(E)} \left(1 + O\left({\omega^2 \over C} \right) \right) \ ,
\end{align}
where $C \gg 1$ is the heat capacity. Corrections in the last equation are vanishing in the thermodynamic limit. Putting it together we have
\begin{align}
&{1\over Z(\beta)} \int dE_1 dE_2 ~ \rho(E_1) \rho(E_2) ~ e^{-(\beta-\tau) E_1 - \tau E_2}
~ \la |{\cal O}_{E_1E_2}|^2\ra \\ 
\approx & {1\over Z(\beta)} \int dE ~ \rho(E) e^{-\beta E} \int_{-\infty}^\infty  d\omega~ e^{\omega (\tau - {\beta \over 2})} \wh{\cal G}_{\beta(E)}(\omega) \\
\approx & \int_{-\infty}^\infty  d\omega~ e^{\omega (\tau - {\beta \over 2})} \wh{\cal G}_{\beta}(\omega) \\
=& {\cal G}_\beta(\tau) \ .
\end{align}
In the second equality, the integral over $E$ is computed by the saddle approximation and sets $\beta = S'(E)$.

\subsection{Four-point function}

Now we compute the four-point correlator \eqref{2ptTMFT}. In the energy basis it turns out to be
\begin{align}\label{4ptETMFT1}
\la {\cal O}_{E_1E_2} {\cal O}_{E_2E_3} {\cal O}_{E_3E_4} {\cal O}_{E_4E_1} \ra = 
&\la |{\cal O}_{E_1 E_2}|^2 \ra \la |{\cal O}_{E_3 E_4}|^2 \ra 
\left( {\delta(E_1 - E_3) \over \rho(E_1)} + 
{\delta(E_2 - E_4) \over \rho(E_2)}\right)
\\
+& \la {\cal O}_{E_1E_2} {\cal O}_{E_2E_3} {\cal O}_{E_3E_4} {\cal O}_{E_4E_1} \ra_c \ ,
\label{4ptETMFT2}
\end{align}
where the disconnected part is determined by \eqref{2ptETMFT}, while the connected part is 
\begin{align}\label{4ptTMFTconn}
\la {\cal O}_{E_1E_2} {\cal O}_{E_2E_3} {\cal O}_{E_3E_4} {\cal O}_{E_4E_1} \ra_c = 
(\wh{\cal G}(\omega_1) \wh{\cal G}(\omega_2) \wh{\cal G}(\omega_3) \wh{\cal G}(\omega_4))^{1/2} ~
{\delta(E_1 + E_3 - E_2 - E_4) \over \rho(E)^3}
\ ,
\end{align}
where 
\begin{align}
E = {1\over 4} \sum_{j=1}^4 E_j \ , \quad \omega_1 = E_1 - E_2 \ , \quad \omega_2 = E_2 - E_3 \ , \quad \omega_3 = E_3 - E_4 \ , \quad \omega_4 = E_4 - E_1 \ .
\end{align}
Let's check that inserting these expressions into \eqref{4ptTMFT} we obtain the TMFT four-point correlator \eqref{4ptTMFTfact}. Each of the three terms in \eqref{4ptETMFT1}, \eqref{4ptETMFT2} correspond to the three terms in \eqref{4ptTMFT} respectively. The idea of the computation is to change coordinates $E_j$ to $E, \omega_j$
\begin{align}
dE_1 dE_2 dE_3 dE_4 &= dE d\omega_1 d\omega_2 d\omega_3 \\
& = \delta(\omega_1 + \omega_2 + \omega_3 + \omega_4) ~ 
 dE d\omega_1 d\omega_2 d\omega_3 d\omega_4 \ .
\end{align}
We also have
\begin{align}
E_1 &= E+ { 2\omega_1 + \omega_2 - \omega_4 \over 4}  \ , \quad 
E_2 = E+  {2\omega_2+\omega_3 - \omega_1 \over 4} \ , \\ 
E_3 &= E+   {2\omega_3+ \omega_4 - \omega_2 \over 4} \ , \quad 
E_4 = E+ {2\omega_4 +\omega_1 - \omega_3 \over 4} \ .
\end{align}
The entropies in $\rho(E_j) = e^{S(E_j)}$ are expanded around the average value $E$ to linear order in $\omega_j$, Further corrections are suppressed in the thermodynamic limit. For example
\begin{align}
\rho(E_1) &\approx e^{S(E) + S'(E)  {2\omega_1 + \omega_2 - \omega_4 \over 4} }  \left( 1+ O\left( \omega_j^2 \over C \right) \right)\ .
\end{align}
The integral over the average energy $E$ will then set $S'(E) = \beta$ as usual. Similarly for other entropic factors.

Putting it together, we have for the first term in \eqref{4ptETMFT1}
\begin{align}
&{1\over Z(\beta)}
\int \prod_{j=1}^4 (dE_j ~ \rho(E_j) e^{-\beta_j E_j})~
\la |{\cal O}_{E_1 E_2}|^2 \ra \la |{\cal O}_{E_3 E_4}|^2 \ra 
 {\delta(E_1 - E_3) \over \rho(E_1)} \\
 \approx&
{1\over Z(\beta)}
\int dE ~ \rho(E) e^{-\beta E} 
\int_{-\infty}^\infty \prod_{j=1}^4 d\omega_j  ~ \delta(\omega_1 + \omega_2)\delta(\omega_3 + \omega_4) 
~ \wh{\cal G}_{\beta(E)}(\omega_1) \wh{\cal G}_{\beta(E)}(\omega_3)
\\
&  
\exp\left\{ -(\beta+\beta_1){ 2\omega_1 + \omega_2 - \omega_4 \over 4}
- \beta_2 {2\omega_2+\omega_3 - \omega_1 \over 4}
- \beta_3 {2\omega_3+ \omega_4 - \omega_2 \over 4}
- \beta_4 {2\omega_4 +\omega_1 - \omega_3 \over 4} 
\right\}
 \\
\approx& 
\int_{-\infty}^\infty d\omega_1 d\omega_3 ~ e^{(\tau_{12} - {\beta \over 2}) \omega_1 + (\tau_{34} - {\beta \over 2}) \omega_3}
~ \wh{\cal G}_{\beta}(\omega_1) \wh{\cal G}_{\beta}(\omega_3) 
\\
=& {\cal G}_\beta(\tau_{12}) {\cal G}_\beta(\tau_{34}) \ .
\end{align}
In the second equality we computed the $E$ integral by saddle approximation and therefore substituted $\beta(E)$ by $\beta$.

The other two terms in \eqref{4ptETMFT1}, \eqref{4ptETMFT2} are computed similarly to above and give the other two terms in \eqref{4ptTMFTfact}. In particular, note that the connected part of the correlator \eqref{4ptTMFTconn} gives the ``crossed'' term ${\cal G}_\beta(\tau_{13}) {\cal G}_\beta(\tau_{24})$.

\subsection{Semi-classical limit of JT gravity}

An example of TMFT arises as a semi-classical limit $G_N \to 0$ of JT gravity coupled to a free scalar field. This is usually called Generalized Free Field (GFF). JT gravity will be discussed in more detail in section \ref{sec.JTmatterMatrixModel}.

The two-point function of GFF, briefly discussed in section \ref{MMforETH}, is
\begin{align}
{\cal G}_\beta(\tau) = \left( {\pi/\beta \over \sin (\pi \tau/  \beta)} \right)^{2\Delta} \ .
\end{align}
The Laplace transform is 
\begin{align}
\wh{\cal G}_\beta(\omega) = 
 {(2\pi/\beta)^{2\Delta -1} \over 2\pi \Gamma(2\Delta)}     \Gamma\left( \Delta \pm {i\beta\over 2\pi} \omega \right) \ .
\end{align}
TMFT correlators with this particular form of the two-point function arise in JT gravity in the semi-classical limit due to\footnote{The reader not familiar with the corresponding results in JT, may wish to return to this after looking at the formulae \eqref{eq.2ptGrav}, \eqref{4pt diag} in section \ref{sec.JTmatterMatrixModel}.}
\begin{align}\label{JT2ptE}
{\Gamma(\Delta \pm i \sqrt{E_1}\pm i \sqrt{E_2}) \over \Gamma(2\Delta)} & \approx 
{1\over \rho(E)}
~ {(2\sqrt{E})^{2\Delta -1} \over 2\pi \Gamma(2\Delta)}     \Gamma\left( \Delta \pm {i\over 2\sqrt{E}} \omega \right) \ ,
\end{align}
where ``$\pm$'' in the LHS means a product of four gamma functions for all choices of signs. We defined $E= {E_1 + E_2 \over 2}, \omega = E_1 - E_2$. The semi-classical limit\footnote{In our conventions we set ${\bar \phi_r \over 4G_N} = 1$, where $\bar \phi_r$ is the boundary value of of the dilaton and has dimensions of length. To restore $G_N$ one simply rescales all energies $E_j \to E_j {\bar \phi_r \over 4G_N}$. Therefore, the semi-classical limit $G_N \to 0$ corresponds to the high-energy limit. } corresponds to $E_1, E_2 \gg |E_1 - E_2|$ and the semi-classical density of states is
\begin{align}
\rho(E) \approx {1\over (2\pi)^2} e^{2\pi \sqrt{E}} \ .
\end{align}
To obtain the semi-classical limit of the four-point and higher correlators in JT gravity we need the limit of the so-called $sl(2,{\mathbb R})$ 6j-symbol. In appendix C of \cite{Jafferis:2022wez} we show that 
\begin{align}\label{6jclass}
 \left\{
\begin{matrix}
\Delta & \sqrt{E_1} & \sqrt{E_2} \\
\Delta & \sqrt{E_3} & \sqrt{E_4}
\end{matrix}
\right\} 
~\approx 
{\delta(E_1 + E_3 - E_2 - E_4) \over \rho(E)} \ , \qquad 
E = {1\over 4} \sum_{j=1}^4 E_j \ .
\end{align}
This is again in the limit $E_j \gg |E_i-E_k|$.

\subsection{Higher TMFT correlators}

Higher correlators in a general TMFT can be also written in the energy basis. It turns out that they are computed by the so-called ``chord diagrams''. Such chord diagrams can also be used to compute JT/matter correlators and are described in detail in section 2 of \cite{Jafferis:2022wez}. The rules for a general TMFT are essentially the same, but instead of the 6j-symbol one uses the RHS of \eqref{6jclass} with an arbitrary density of states $\rho(E)$ and instead of the JT two-point function in the LHS of \eqref{JT2ptE} one uses \eqref{2ptETMFT}.

\subsection{TMFT Matrix Model}

The semi-classical limit of the JT/matter matrix model potential \eqref{eq:constraintsquared}, which we generalize to a TMFT with an arbitrary two-point function $\wh{\cal G}_\beta(\omega)$ and an arbitrary density of states $\rho(E) = e^{S(E)}$, is
\begin{align}\label{MMV}
V(R) &= N\left( {1\over 2}\sum_{ab} g_{ab} |R_{ab}|^2 + {1\over 4} \sum_{abcd}g_{abcd} R_{ab} R_{bc} R_{cd} R_{da} \right) \ , \\
g_{ab} & 
%=
%-1 + {1\over 2} {\rho(\bar s + {3s_{ab}\over 2}) \over \rho(\bar s+{s_{ab}\over 2})}
%+ {1\over 2} {\rho(\bar s - {3s_{ab}\over 2}) \over \rho(\bar s-{s_{ab}\over 2})}\\
= 2\sinh^2  {\beta({\overline E})\omega \over 2}  \ , \\
g_{abcd} & = {1\over 2}{\delta(E_a - E_c) \over \rho(E_a)} +  {1\over 2} {\delta(E_b - E_d) \over \rho(E_b)} - {\delta(E_a + E_c - E_b-E_d) \over \rho(E)} \ ,
\end{align}
where $\overline E = {E_a + E_b \over 2}, \ \omega = E_a - E_b, \ E = {1\over 4}(E_a + E_b+E_c + E_d)$, and $\beta(\overline E) = S'(\overline E)$ is the inverse temperature corresponding to the energy $\overline E$.

The matrix $R_{ab}$ is a rescaling of ${\cal O}_{ab}$
\begin{align}
{\cal O}_{ab}:=%{1\over  \sqrt{\rho(\overline E) }} ~ 
%(2\bar s)^{\Delta} 
%\left(  {\Gamma(\Delta \pm i s_{ab}) \over 2\pi \Gamma(2\Delta)}  \right)^{1/2} 
\left(\wh{\cal G}_{\beta(\overline E)}(\omega) \over \rho(\overline E) \right)^{1/2}
R_{ab} \ ,
\end{align}
where $\overline E = {E_a + E_b \over 2}, \omega = E_a - E_b$. With this rescaling the exact (planar) two-point function is $\la |R_{ab}|^2 \ra_{\disk} = 1$, see \eqref{2ptETMFT}.

The matrix model description of thermal mean field theory follows from the more general discussion of JT gravity with matter. In particular, the so called ``unlacing rules'' of the 6j-symbol that are crucial in the analysis of \cite{Jafferis:2022wez} are satisfied by the RHS of \eqref{6jclass}. It might be interesting to repeat the analysis of \cite{Jafferis:2022wez} directly in the semi-classical limit.

This matrix model describes TMFT in the thermodynamic limit. In this limit, we zoom into the window of energies in the middle of the spectrum. This is analogous to the double-scaling limit in the matrix model dual of JT gravity\footnote{And more generally, matrix model duals of Minimal String Theory.} discussed in the next section.

\section{Example 2: JT/matter matrix model \label{sec.JTmatterMatrixModel}}

\subsection{Matter correlation functions in JT gravity\label{sec.diskCorrelators}}

Let us now turn to our main application of the structure we defined above, namely the matrix model description of JT gravity coupled to a scalar field. The theory is defined via the action
\begin{equation}
S_{\rm JT} = -S_0 \chi -\frac{1}{2} \int_{\cal M}\sqrt{g}\phi \left( R + 2 \right)  - \int_{\partial {\cal M}} \sqrt{h} \phi \left(K-1 \right) + S_{\rm m},
\end{equation}
with the matter action
\begin{equation}
S_{\rm m} = \frac{1}{2}\int_{\cal M}\sqrt{g} \left( g^{ab} \partial_a \varphi \partial_b \varphi + m^2 \varphi^2  \right)\,.
\end{equation}
Here $\chi$ is the Euler number of the two-manifold ${\cal M}$ over which the Lagrangian density is integrated, $\phi$ is the dilaton field which forms part of the definition of the two-dimensional gravity theory under study, while $\varphi$ is an additional matter scalar field of mass $m$. The term integrated over the boundary of the manifold $\partial {\cal M}$ contains a Gibbons-Hawking-York term in the form of the integrated extrinsic curvature $K$ as well as a boundary cosmological constant. Without the addition of the matter scalar $\varphi$ this theory has been shown to be equivalent to a matrix model \cite{Saad:2019lba}, while here (see also the longer companion paper \cite{Jafferis:2022wez}) we generalize this to the theory including the matter field.

We start by laying out a few useful facts and computations in the theory above. All expressions quoted here can be obtained by using a convenient set of Feynman rules, developed in \cite{Mertens:2017mtv} and reviewed in \cite{Jafferis:2022wez}. Firstly, the inclusion of the Euler number in the action means that amplitudes have an expansion in the topology of the manifold ${\cal M}$ which is used to calculate them gravitationally.

We start with the leading-order, where the topology is that of a disk. At disk level, the density of states of JT gravity is given by 
\begin{equation}\label{eq.JTdiskDensity}
\rho_0(E)dE = \frac{1}{2\pi^2}\sinh \left( 2\pi \sqrt{E} \right)dE = \frac{1}{\pi^2}  s \sinh \left( 2\pi s \right)ds\,,
\end{equation}
where we have defined the variable $s^2 = E$.

\subsubsection*{Disk correlation functions}

Turning now to correlation functions\footnote{More details may be found in the original references studying disk correlators of the Schwarzian, \cite{BAGRETS2016191,Mertens:2017mtv, Yang:2018gdb, Kitaev:2018wpr,Suh:2020lco}.}, we focus on the two-point function, which at disk level reads
\begin{equation}\label{eq.2ptGrav}
    \la \tr e^{-\beta H} \calo(\tau) \calo(0) \ra_{\disk}
    =   e^{S_0} \int_0^\infty ds_1 ds_2~
    \rho(s_1) \rho(s_2) 
    ~ e^{-(\beta_1 s_1^2 + \beta_2 s_2^2)} 
    \Gamma_{12}^\Delta
    \,.
\end{equation}
Here $\beta_1 = \tau, \beta_2 = \beta - \tau$ and we have also introduced the short-hand notation 
\begin{equation}
\Gamma_{12}^\Delta : = \frac{\Gamma(\Delta \pm i s_1 \pm i s_2)}{\Gamma(2\Delta)} = \Gamma(2\Delta)^{-1}\Gamma(\Delta +i s_1 +is_2)\Gamma(\Delta +i s_1 -is_2)\Gamma(\Delta -i s_1 +is_2)\Gamma(\Delta -i s_1 -is_2)\,.
\end{equation}
Using this, we can express the four-point function as
\begin{align}
\label{4pt diag}
\la \tr e^{-\beta H} \calo(\tau_1) \dots \calo(\tau_4) \ra_{\disk} &= e^{S_0} \int_0^\infty 
     \prod_{i=1}^4 \left( ds_i ~\rho(s_i) e^{-\beta_i s_i^2} \right)
     {\left( \Gamma_{12}^\Delta \Gamma_{23}^\Delta \Gamma_{34}^\Delta \Gamma_{41}^\Delta \right)^{1/2}  } \nonumber\\
    &\qquad \qquad \times \left( 
     {\delta(s_1 -s_3) \over \rho(s_1)} 
     + {\delta(s_2 -s_4) \over \rho(s_2)}
     + \left\{\begin{array}{ccc}
	 		\Delta & s_1 & s_2 \\
	 		\Delta & s_3 & s_4
	 	\end{array}\right\} 
     \right)
    \ ,
\end{align}
The third term in this expression uses the bracket notation for the 6j-symbol of the $\mathfrak{sl}(2,\mathbb{R})$ algebra. Expressions for any higher-point functions can be obtained (see \cite{Mertens:2017mtv,Jafferis:2022wez}), but the two- and four-point functions reviewed above shall suffice for present purposes. We will, however, need to consider correlation functions on geometries of different topology, starting with the so-called `double-trumpet' \cite{Saad:2019lba}, which is topologically a cylinder.

\subsubsection*{Direct match to JT + matter theory}
Let us proceed somewhat naively to determine the couplings of the ETH matrix model describing matter-coupled JT gravity. In fact, the answers we arrive at are correct, but more machinery is needed to justify them properly, such as an understanding of how to obtain them in a double-scaling limit. This and other issues will be discussed in the remainder of this paper, including studying the theory on higher topologies, i.e. away from the disk level.

 We start by matching the disk density of states
\begin{equation}\label{eq.MatchDensity}
\left\langle {\rm Tr} e^{-\beta H} \right\rangle = e^{S_0} \rho_0(E) = e^{S_0} \sinh\left(2\pi \sqrt{E} \right)\,,
\end{equation}
which involves choosing the potential $V$ in \eqref{eq.DysonBetaMeasure} appropriately. Computing the full disk two-point function in the matrix model, as in \eqref{eq.ETHMatrixModelTwoPoint}, leads to the identification
\begin{equation}\label{eq.MatchTwoPoint}
\overline{{\cal O}_{ab} {\cal O}_{ba}} = e^{-S_0} \frac{\Gamma(\Delta \pm i \sqrt{E_a } \pm i \sqrt{E_b})}{\Gamma(2\Delta)} := e^{-S_0} \Gamma^\Delta_{ab}.
\end{equation}
In an analogous fashion, one can determine the quartic non-Gaussianity of the ETH matrix model for JT + matter. The computation is performed in \cite{Jafferis:2022wez} and leads to the identification
\begin{equation}\label{eq.MatchFourPoint}
e^{-3 S(\bar{E})} g^{(4)}_{\calo}(E_a,E_b,E_c,E_d) = e^{-3 S_0} (\Gamma^\Delta_{ab} \Gamma^\Delta_{bc} \Gamma^\Delta_{cd} \Gamma^\Delta_{da})^{1/2} \left\{
\begin{matrix}
\Delta & s_a & s_b \\
\Delta & s_c & s_d
\end{matrix}
\right\}.
\end{equation}
In the companion paper we describe a systematic procedure determining the full ${\cal O}$ potential in an expansion working in the number of `bulk' line crossings.

\subsubsection*{Double-trumpet correlation functions}
We now include the scalar one-loop determinant in the otherwise `empty' double trumpet, that is we consider the effect of the scalar field on the trumpet absent any explicit insertions of the ${\cal O}$ operator at the boundary,
\begin{align}	\label{eq:dttoy}
\braket{\tr e^{-\beta_L H} \tr e^{-\beta_R H}}
	&= \int_0^\infty db ~b \, Z_{\text{tr}}(\beta_L,b) Z_{\text{tr}}(\beta_R,b) \, Z_{\text{scalar}}(b)\,,
\end{align}
that is it differs from the usual JT double trumpet only by the inclusion of $Z_{\text{scalar}}(b)$. The factors $Z_{\text{tr}}(\beta_R,b)$ are given by the trumpet partition function for a geometry characterized by an asymptotically AdS$_2$ boundary of length $\beta$ and a geodesic boundary of length $b$ in the interior. To obtain the explicit expression $Z_{\text{tr}}(\beta_R,b) = \frac{1}{2\sqrt{\pi \beta}}  e^{-\frac{b^2}{4\beta}}$ one integrates over the fluctuations of the asymptotic boundary weighted by the Schwarzian measure \cite{Saad:2019lba} giving the one-loop exact result \cite{Stanford:2017thb} we quoted. In the companion paper, we show how to evaluate the scalar determinant in several different ways, useful for various different points of view. One can show that

\begin{align}
	Z_{\text{scalar}}(b) &=  \sum_{n = 0}^\infty \frac{e^{- n \Delta b}}{(1-e^{-b})(1-e^{-2b}) \dots (1-e^{-nb})} = 1+\sum_{\Delta^\prime \in \mathcal{S}} \frac{e^{- \Delta^\prime b}}{1 - e^{- b}}\label{eq:second}
	\\ &= \exp\left(\sum_{w = 1}^\infty \frac{e^{- w b \Delta }}{w(1 - e^{- w b})}\right)\label{eq:third}\,.
\end{align}
One important feature, apparent in all representations, is the presence of a UV divergence for $b\rightarrow 0$. This divergence is reproduced from our matrix model, in the double-scaling limit.  The first expression above, makes apparent the relation to the spectrum of conformal primary operators and their descedants, for example the $n=1$ term can be recognized as a sum over the primary state of dimension $\Delta$ together with all of its descendants. Similary, the $n=2$ term sums over the double trace operators with dimensions $2\Delta + 2m$, together with all descendants. Analogous interpretations continue to hold for all higher $n$. In the last expression in \eqref{eq:second}, $\cal{S}$ is defined to be the set of scaling dimensions of all primary operators in the bosonic generalized free field (other than the identity). The $\frac{1}{1-e^{-b}}$ factor may be expanded in a geometric series, which is associated to the descendants. The second formula is naturally related to the Selberg trace formula for the heat kernel of a scalar operator on the double trumpet, that is as a sum over  (multiply wound) primitive geodesics on this particular hyperbolic manifold.

\subsubsection*{Double-trumpet  two-point function}
We will next determine the two-point correlation function of the JT-matter theory, with an ${\cal O}$ insertion on each boundary. In the companion paper (see Appendix D of \cite{Jafferis:2022wez}) it is shown that the result arranges itself into a sum over terms corresponding to the structure present in the expansion  \eqref{eq:second} above.  Focusing explicitly on the first three such contributions, we obtain
\begin{align}
&\braket{\tr e^{-\beta_L H} \calo \tr e^{-\beta_R H} \calo}_{\cyl} \\
&= 
\int_0^\infty ds_L ds_R ~\rho(s_L)\rho(s_R) 
e^{- (\beta_L s_L^2 + \beta_R s_R^2) }
\left(\Gamma^\Delta_{LL} \Gamma^{\Delta}_{RR}
\right)^{1/2}
\\
&\times  
\left(
{\delta(s_L - s_R) \over \rho(s_L)} 
+ 
\begin{Bmatrix}
   \Delta & s_L & s_R \\
   \Delta & s_R & s_L
\end{Bmatrix}
+ 
\sum_{m=0}^\infty  
\begin{Bmatrix}
   2\Delta + 2m & s_L & s_R \\
   \Delta & s_R & s_L
\end{Bmatrix}
+ \dots 
\right) \,,
\label{2pt-dt2}
\\
&:=
\raisebox{-.3in}
{\includegraphics[scale=.15]{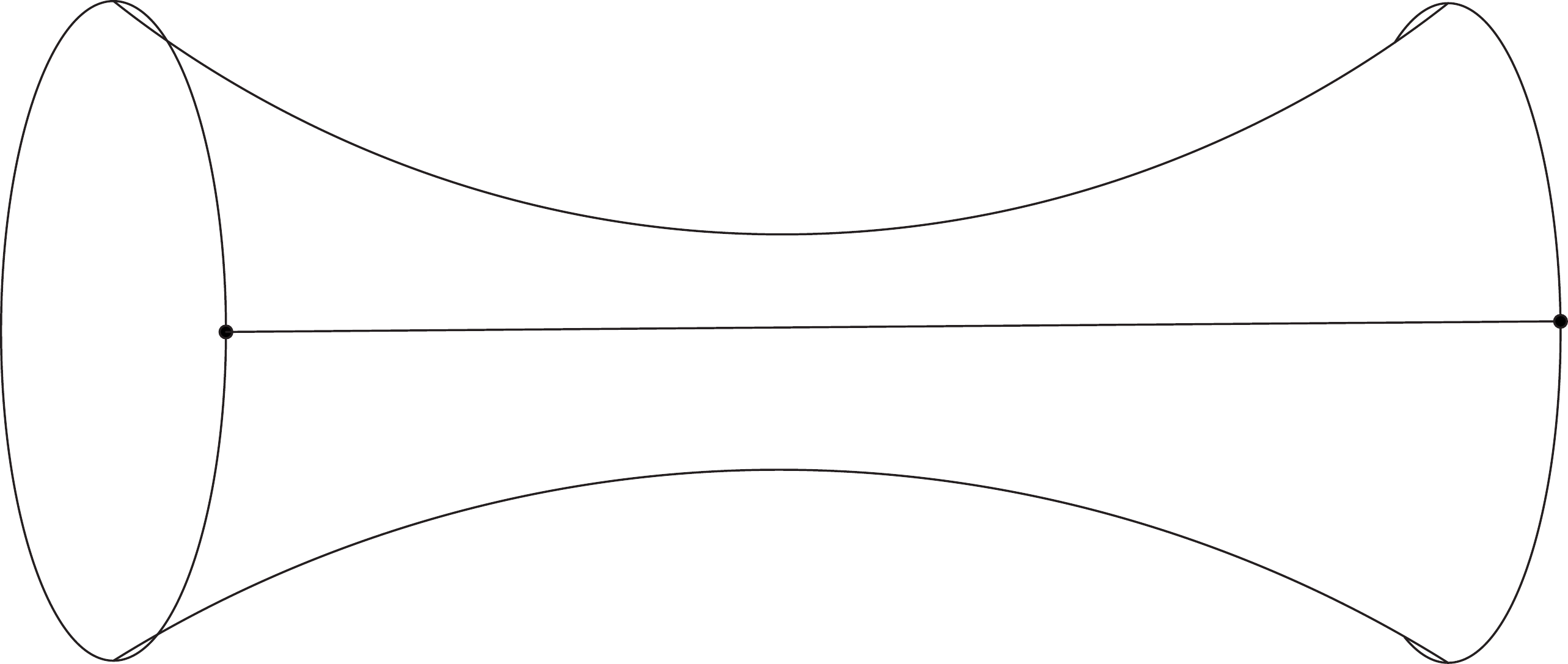} }
+ 
\raisebox{-.35in}
{\includegraphics[scale=.15]{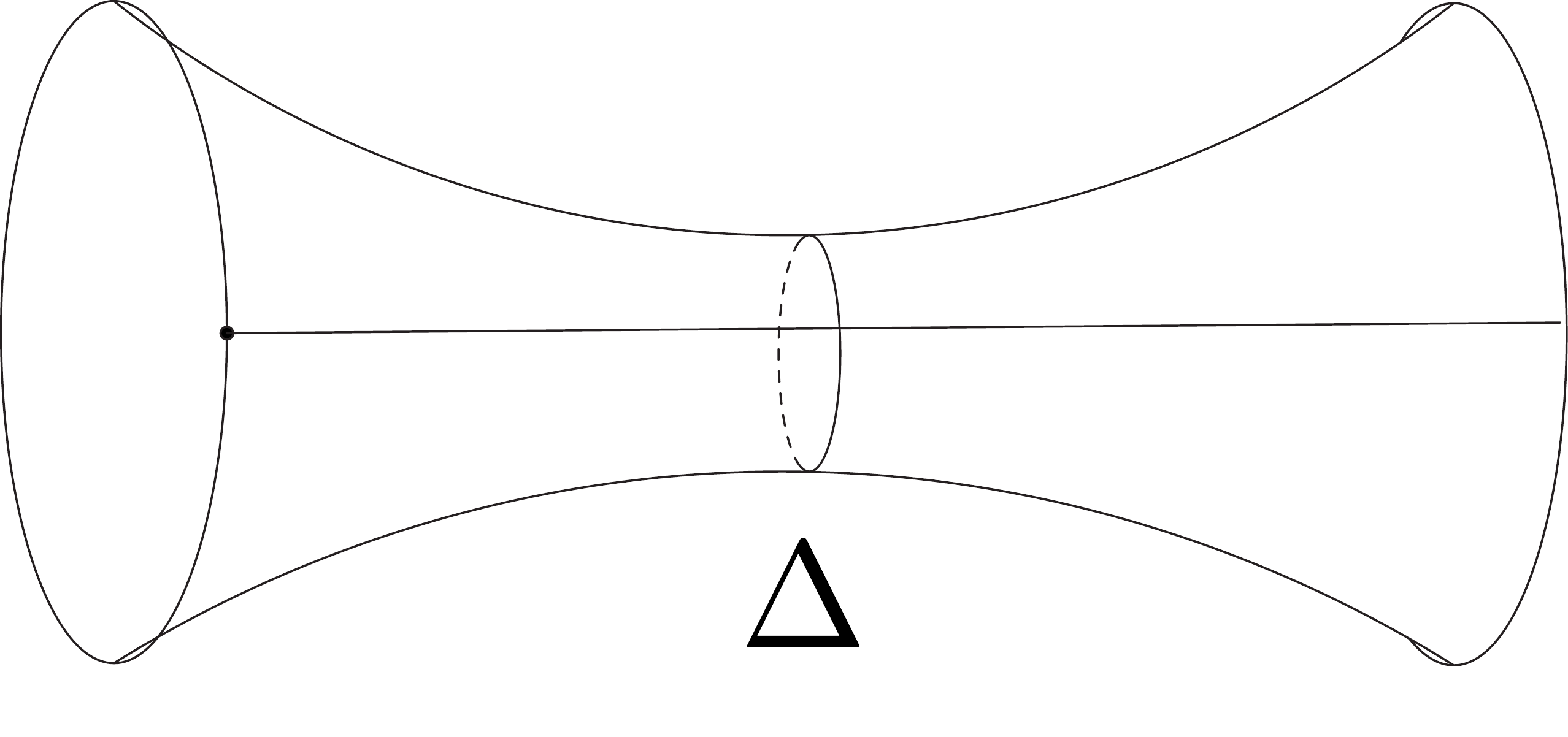} }
+ \sum_{m=0}^\infty 
\raisebox{-.35in}
{\includegraphics[scale=.15]{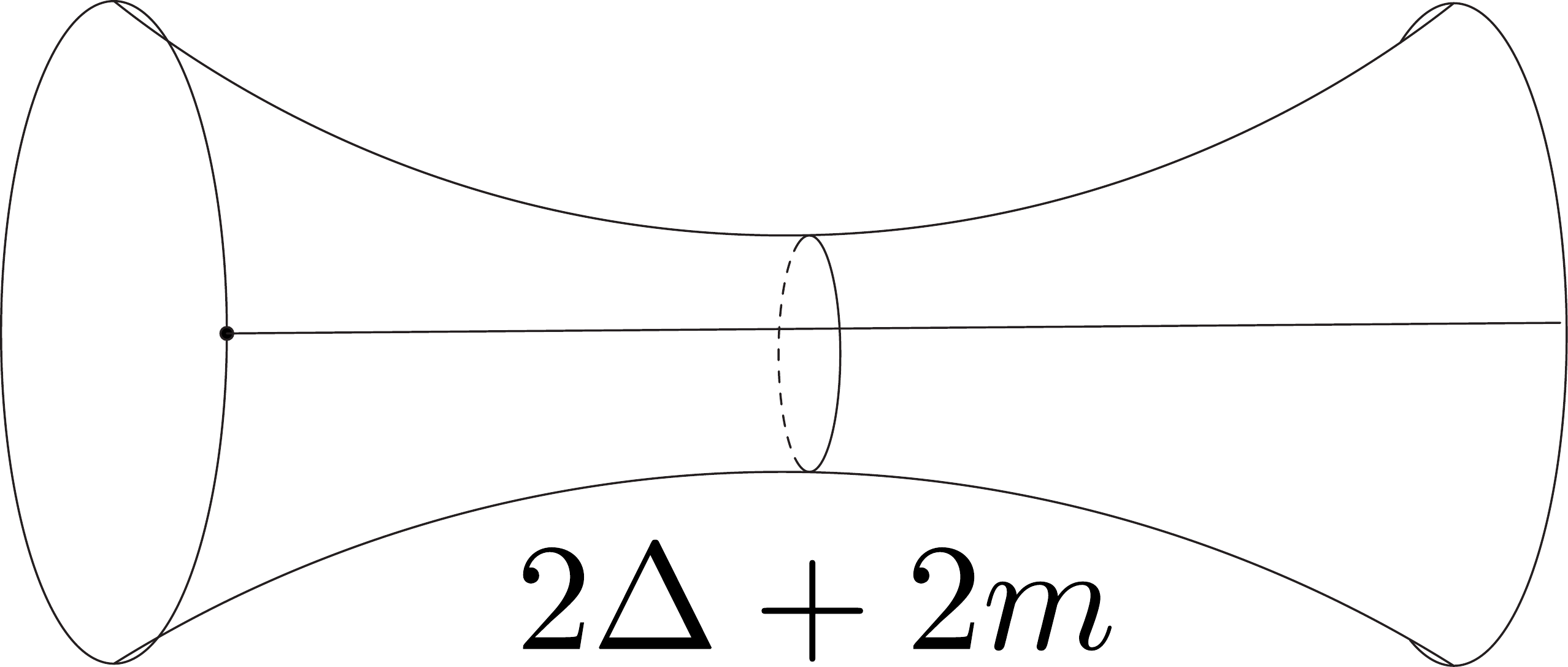} }
+ \dots \ .
\end{align}
where $\Gamma_{LL}^\Delta = \frac{\Gamma(\Delta)^2}{\Gamma(2\Delta)} \Gamma(\Delta + is_L) \Gamma(\Delta - is_L)$, and analogously for $\Gamma^\Delta_{RR}$ in terms of $s_R$.
 In the expression above, the first term results from a simple propagation of an ${\cal O}$ line along a geodesic connecting the two boundary ${\cal O}$ insertions through the bulk. The second takes the form of a geodesic ${\cal O}$ propagation through the bulk, this time crossed by an ${\cal O}$  bulk loop along a closed geodesic, the intersection of boundary-to-boundary geodesic and bulk loop giving a rise to the 6j-symbol labelled by the scaling dimension $\Delta$, as shown. Finally the third term shows the analogous process, but now involving  the sum over double-traces of dimension $2\Delta + 2m$ in the loop. The higher contributions we omitted correspond in a similar fashion to the $n\ge 3$ terms in the expansion of the scalar one-loop determinant \eqref{eq:second} above.

\subsubsection*{Higher-genus correlation functions}
The general higher-genus, $n$-boundary geometry can be assembled via a pants decomposition of a genus$-g$ Riemann-surface with $n$ geodesic boundaries, glued to $n$ trumpet geometries. In the companion paper we accumulate evidence that an arbitrary ${\cal O}$ $n-$point correlation function at genus $g$ can be computed as follows.

Let us first ignore the contribution to the correlator from the determinant of the scalar field. Then the correlator is computed by summing over all geodesics that connect the $\calo$ operators on the asymptotic boundaries, with a weighting of $e^{- \Delta \ell}$ for each geodesic, where $\ell$ is a renormalized length. These geodesic configurations are classified by their topologies. Each topology is represented by a set of lines drawn on the surface with no voluntary intersections. These lines divide the surface into subregions, and each subregion is characterized by the number of boundaries $n$ and the genus $g$. Each boundary of each subregion is labeled by an $s$ parameter. For every subregion, we should include a factor of $\braket{\rho(s_1) \ldots \rho(s_n)}_{g,n}$, which we define to be the inverse Laplace transform of $Z_{g,n}(\beta_1,\ldots,\beta_n)$, which is the path integral defined in equation (127) of \cite{Saad:2019lba}. For every intersection of two bulk lines, we include a factor of a 6j symbol that depends on the four adjacent $s$ parameters. An asymptotic AdS boundary with Euclidean length $\beta$ is assigned a factor of $e^{-\beta s^2}$. Finally, one should integrate over all of the $s$ parameters from $0$ to $\infty$. For disk topologies, these Feynman rules become the rules of \cite{Mertens:2017mtv}, which we have reviewed in \cite{Jafferis:2022wez}.

To include the contribution from the scalar field matter determinant, we also should sum over all ways of drawing closed geodesics on the spacetime. The rules for an intersection of two geodesics are still as above. The scaling dimensions assigned to the closed geodesics should take all values in $\cal{S}$, which was defined in \eqref{eq:second}. Finally, wherever there is a simple closed geodesic, we include the following function of its two adjacent $s$ parameters:
\begin{equation}
	\int_0^\infty db ~ b ~ {e^{-\Delta' b} \over 1-e^{-b}} ~ {2\over b} \cos(b s_1) ~ {2\over b} \cos(b s_2),
\end{equation}
where $\Delta'$ is the dimension of the primary operator propagating on the closed geodesic. This ensures that the closed geodesic is weighted by ${e^{-\Delta' b} \over 1-e^{-b}}$ in the moduli space integral.

The evidence we have for the above conjecture comes from our computation of the double-trumpet two-point function as well as a computation on the pair of pants, presented in \cite{Jafferis:2022wez}. See also \cite{Saad:2019pqd, Iliesiu:2021ari, Blommaert:2020seb}, which discuss the two-point function on the disk with a handle.

\subsection{Matching to the ETH matrix model}
We now have assembled all the data needed to fully specify the free coupling functions of the general ETH matrix model. This can be viewed alternatively as an exercise demonstrating the usefulness of the general ETH matrix model, and the generalized ETH, or as an extension of the JT matrix model \cite{Saad:2019lba} to include a scalar field. The task at hand is to specify the higher-order coupling functions $G^{(n)}(E_a, \ldots E_n)$ in the ETH matrix model \eqref{eq.ETHTwoMatrixModel}. We may achieve this in two different ways\footnote{We refer here to deducing directly the double-scaled matrix model. In the longer companion paper we also present two classes of regulated matrix models, which recover the results here in the double-scaled limit.}. 
\begin{itemize}
\item We use the nearly-conformal invariance and locality of the JT-matter theory in order to constrain the potential non-linearities directly. The leading non-linear couplings of the ETH matrix model are given as a 'constraint-squared' type potential. Roughly speaking we translate the observation around Equation \eqref{eq:rightGFFcorrelator} into the constraint 
\begin{equation}
			[\calo(t_1),\calo(t_2)] =\left\{
			\begin{array}{cc}
	-\frac{2 i \sin(\pi \Delta)}{|t_1 - t_2|^{2 \Delta}} \, \text{sign } (t_1 - t_2), & \Delta \notin \mathbb{Z}_{\ge 1} \\
	\frac{2 \pi i}{(2 \Delta - 1)!} (-1)^{\Delta -1} \delta^{(2 \Delta - 1)}(t_{12}), & \Delta \in \mathbb{Z}_{\ge 1}
\end{array}\right. .
\label{eq:GFFconstraint}
\end{equation}
This constraint is saying the the commutator is proportional to a c-number, or in other words the unit operator.
This constraint finds its representation in the matrix model as a specific non-Gaussianity, as we shall see.
\item We deduce the higher-order couplings by matching correlation functions to the gravity predictions at disk level (determined in Section \ref{sec.diskCorrelators}), and conjecture that the matrix model so-defined correctly predicts all higher-genus contributions. This is the analogue of the statement that in the single-matrix model of \cite{Saad:2019lba}, the disk density of states determines the full matrix potential and thus the higher-genus amplitudes. In particular, we match away from the double-scaling limit to a suitably $q-$deformed version of the gravity amplitudes, before taking the $q \rightarrow 1$ limit that recovers the original amplitudes.
\item In addition, whichever strategy we use in order to determine the ${\cal O}$ potential, we need to add counter terms to the $H$ potential, in order to (re-)adjust the disk density of states $\rho_0(E)$ to the desired form -- for example the $\sinh(2\pi \sqrt{E})$ behavior of JT gravity with matter.  That this is needed can already be seen at the Gaussian level for the ${\cal O}$ statistics: in this case we can directly integrate out the ${\cal O}$ matrix to obtain 
\begin{equation}
{\cal Z}_{\rm ETH} = \int dH e^{-V_{\rm SSS}(H) - \tilde V(H)}\,,
\end{equation}
where we denoted the single-trace $H$ potential of \cite{Saad:2019lba} by the initials of its three authors, and 
\begin{equation}
\tilde V(H) = \sum_a V_{\rm ct} (E_a) + \frac{1}{2}\sum_{a,b} \log F(E_a,E_b)\,,
\end{equation}
written in the energy eigenbasis. Here $V_{\rm ct}$ is the counterterm potential we are trying to determine, and $F(E_a,E_b)$ is the coefficient of the quadratic term in the ${\cal O}$ matrix. It is clear that integrating out has changed the naive disk density of states, which we can compensate by a judicial choice of $V_{\rm ct}(H)$. An important result of \cite{Jafferis:2022wez} is that this can be achieved with a single-trace counterterm.
\end{itemize}
Having outlined the general procedure, let us now expand on the two approaches to determine the ${\cal O}$ potential in some more detail.

\subsection{Double-scaling limit}
While we did not specify this explicitly, in a typical application the ETH ansatz \eqref{eq.GaussianETH} is invoked for random matrices $R_{ab}$ of finite dimension, and thus for a locally finite Hilbert space, so that $e^{S(E)} < \infty$.  In fact, for the main application we have in mind, namely 2D gravity, we are interested in the case where the number of eigenvalues, that is the dimension of Hilbert space is scaled to infinity. As has been studied extensively in past applications of matrix models to the theory of 2D gravity (see e.g. the review \cite{Ginsparg:1993is}), in order to pass to the limit of smooth fluctuating surfaces, this must be accompanied by a rescaling of one or several coupling parameters, leaving appropriate ratios finite. In fact, in matching matrix-model correlation functions to the JT+matter (as well as in the semi-classical limit, i.e. thermal mean field theory discussed above) expressions, we implicitly assume that such a double scaling limit has been taken. This double-scaling procedure leaves a theory with a continuous spectrum of eigenvalues supported on a non-compact cut in the complex energy plane, which can be matched to that of JT gravity (\eqref{eq.JTdiskDensity}, below) for $E \in [0, \infty)$. While the direct-matching procedure may therefore seem a bit ad-hoc at first, we in fact establish these results more carefully by introducing two classes of finite regulated matrix models, the so-called `$q-$deformed' and `Selberg models'. Both of these involve introducing a regulator, which renders the Hilbert-space dimension finite for $q<1$ and is chosen so that in the limit $q\rightarrow 1$, we recover the gravity correlation functions, i.e. the matching \eqref{eq.MatchDensity} - \eqref{eq.MatchFourPoint}.  In all cases we fix the (two-)matrix model using only disk data and then proceed to show that it continues to correctly capture topologically non-trivial correlators of JT+matter. We return to the regulated models and their double scaling limits in Section \ref{sec.DoubleScaleMatch} below after discussing the gravitational correlators both at disk level and higher genus that our ETH-matrix model for JT+matter is designed to reproduce.

\subsection{Constraint-squared potential}

In this section we argue that by imposing a constraint on the matrices $H$ and $\calo$, we can construct a matrix model that correctly computes the disk correlators of JT gravity minimally coupled to a scalar field. Our argument is not rigorous, but it is modeled after a rigorous result in 1D CFT, which states that if $\calo$ is a primary with dimension $\Delta$ and if the spectrum of primary operators appearing in the $\calo \calo$ OPE is that of a bosonic generalized free field (GFF) with dimension $\Delta$, then all of the correlators of $\calo$ are exactly those of the GFF.\footnote{See \cite{Jafferis:2022wez} for the proof. We thank Dalimil Maz\'{a}\v{c} for his invaluable assistance.} JT gravity with matter on the disk admits a semiclassical limit in which the correlators become those of a GFF, and one might expect that even away from the semiclassical limit, there is some condition that can be placed on the operators $\calo$ and $H$ that is only obeyed by gravitational correlators computed using the Feynman rules discussed towards the end of section \ref{sec.diskCorrelators}. If such a condition exists, we may deduce it as follows. First, note that the $\calo \calo$ OPE in the GFF takes the form
\begin{equation}
	\label{eq:OPE}
	\calo(\tau) \calo(0) = \frac{1}{\tau^{2 \Delta}} + \sum_{n = 0}^\infty \tau^{2 n} [\calo \calo]_n + \text{descendants}.
\end{equation}
Aside from the identity, the primary operators above have dimensions $2 \Delta + 2n$ for $n$ a nonnegative integer. Equation \eqref{eq:OPE} implies \eqref{eq:GFFconstraint}, which we repeat here: 
\begin{equation*}
	[\calo(t_1),\calo(t_2)] = \lim_{\epsilon \rightarrow 0} \left[ \frac{1}{( i t_{12} + \epsilon)^{2 \Delta}} - \frac{1}{( i t_{12} - \epsilon  )^{2 \Delta}}\right] 
	= \left\{
	\begin{array}{cc}
		-\frac{2 i \sin(\pi \Delta)}{|t_1 - t_2|^{2 \Delta}} \, \text{sign } (t_1 - t_2), & \Delta \notin \mathbb{Z}_{\ge 1} \\
		\frac{2 \pi i}{(2 \Delta - 1)!} (-1)^{\Delta -1} \delta^{(2 \Delta - 1)}(t_{12}), & \Delta \in \mathbb{Z}_{\ge 1}
	\end{array}\right. \, .
\end{equation*}
In a 1D CFT, \eqref{eq:GFFconstraint} also implies \eqref{eq:OPE}, because any contribution to \eqref{eq:OPE} with a different power of $\tau$ would make an additional nontrivial contribution to the right side of \eqref{eq:GFFconstraint}. Hence, a quadratic operator equation for $\calo$ is enough to guarantee that the spectrum of primaries appearing in the $\calo \calo$ OPE is that of a GFF, which in turn guarantees that the correlators of $\calo$ are those of a GFF. We emphasize that the proof of this assumes the usual bootstrap axioms of conformal invariance and OPE associativity.

Next, we consider JT gravity away from the semiclassical limit and search for an operator equation for $\calo$ and $H$ that is consistent with the Feynman rules described in section \ref{sec.diskCorrelators} and is quadratic in $\calo$. We could only find one such operator equation:\footnote{If our conjecture in section \ref{sec.diskCorrelators} for the Feynman rules at higher genus is correct, then higher genus correlators also obey this constraint.}
\begin{equation}
	e^{-S_0} \sum_{b}    \left\{\begin{array}{ccc}
		\Delta & s_a & s_b \\
		\Delta & s_c & s_d
	\end{array}\right\} \left[\frac{\calo_{a b} \calo_{bc}}{e^{-S_0} \sqrt{\Gamma_{ab}^\Delta \Gamma_{bc}^\Delta}} - {\delta(s_a-s_c) \over e^{S_0}\rho(s_a)} \right] = \left[\frac{\calo_{a d} \calo_{dc}}{e^{-S_0}\sqrt{\Gamma_{ad}^\Delta \Gamma_{dc}^\Delta}} - {\delta(s_a-s_c) \over e^{S_0}\rho(s_a)} \right]. \label{eq:constraint}
\end{equation}
If we insert the right side of \eqref{eq:constraint} into any correlator, the Feynman rules of section \ref{sec.diskCorrelators} dictate that we should omit diagrams where a bulk line connects the two adjacent $\calo$ insertions. If we insert the left side of \eqref{eq:constraint} into a correlator, the diagrams that contribute are the same, except with an additional crossing of the two bulk lines that end on the adjacent $\calo$ insertions (this means that the two lines could intersect twice. Such a double-crossing can then be undone due to an orthogonality relation of the 6j symbols).

We believe that \eqref{eq:constraint} is the appropriate generalization of \eqref{eq:GFFconstraint} to nearly-conformal CFTs. Due to the constraining power of \eqref{eq:GFFconstraint} in 1D CFTs, we expect that \eqref{eq:constraint} is also highly constraining. Recall that the aforementioned rigorous 1D CFT result rests on three assumptions: the operator equation \eqref{eq:GFFconstraint}, associativity of the OPE, and conformal invariance. We explained above that \eqref{eq:GFFconstraint} should generalize to \eqref{eq:constraint} in a nearly-conformal CFT, and thus we impose \eqref{eq:constraint} as a constraint in a matrix model. If we define
\begin{equation}
	M_{ac}^d := \frac{1}{2}\sum_{b}    
	\left(
	e^{-S_0} 
	\left\{\begin{array}{ccc}
		\Delta & s_a & s_b \\
		\Delta & s_c & s_d
	\end{array}\right\} - {\delta(s_b-s_d) \over e^{S_0}\rho(s_b)}
	\right) 
	\left(
	\frac{\calo_{a b} \calo_{bc}}{e^{-S_0} \sqrt{\Gamma_{ab}^\Delta \Gamma_{bc}^\Delta}} - {\delta(s_a-s_c) \over e^{S_0}\rho(s_a)}
	\right),
\end{equation}
then the ensemble is defined by the following matrix integral:
\begin{equation}
	\int dH d\calo \, \exp\left(- \tr V(H) - \frac{\Lambda}{2} \sum_{a,c,d}|M_{ac}^d|^2\right),
	\label{eq:constraintsquared}
\end{equation}
where $\Lambda$ is a large parameter that enforces $M_{ac}^d = 0$, or \eqref{eq:constraint}, as a constraint. Associativity of the OPE is guaranteed by the fact that we are representing the operators using matrices, and matrix multiplication is associative. Of course, we do not expect this matrix model to respect conformal invariance, but it should reproduce a nearly-conformally invariant theory due to the use of 6j symbols in constructing the matrix potential. In \cite{Jafferis:2022wez} we show that many Schwinger-Dyson equations of this matrix model are solved by the desired gravitational disk correlators, and we also discuss how \eqref{eq:constraintsquared} may be defined away from the strict double-scaling limit. Note that the matrix potential of the $\calo$ integral, which is responsible for the correlations of the generalized ETH ansatz, is bounded from below. The potential $V(H)$ is chosen such that the eigenvalue distribution of $H$ to leading order in $e^{S_0}$ matches the disk density of states of JT gravity.

The question of higher-genus correlators in double-scaled matrix models dual to JT gravity with matter is subtle, and we explore it in two separately-defined models which we introduce below.

\subsection{An iterative procedure to determine the matrix model\label{sec.DoubleScaleMatch}}

We now discuss a matching calculation that allows one to determine the choices of the coupling functions $G^{(n)}$, introduced in \eqref{eq.ETHTwoMatrixModel}, that result in the matrix model computing the correct disk correlators. A careful treatment requires us to regulate the gravitational disk amplitudes such that the energies (or equivalently the $s$ parameters) are integrated over a finite range. This is because loop `t Hooft diagrams in the matrix model (such as on the right hand side of \eqref{eq:interactingtwopointthooft}) involve energy integrals, and a finite spectrum guarantees that these integrals converge. To achieve this, we note that the special functions appearing in section \ref{sec.diskCorrelators} admit $q$-deformations,
\begin{align}
	\rho(s) \rightarrow \rho_q(s) \ , \qquad  \Gamma_{12}^\Delta\rightarrow\Gamma_{12,q}^\Delta\,, \qquad
	\left\{\begin{array}{ccc}
		\Delta_1 & s_1 & s_2 \\
		\Delta_2 & s_3 & s_4
	\end{array}\right\}  
	\rightarrow  
	\left\{\begin{array}{ccc}
		\Delta_1 & s_1 & s_2 \\
		\Delta_2 & s_3 & s_4
	\end{array}\right\}_q \ ,  
\end{align}
in such a way that the usual JT+matter rules are recovered in the limit $q\rightarrow 1$. Precise definitions are provided in \cite{Jafferis:2022wez}. The important point is that the $s$ parameter is integrated only in the range $(0,\frac{\pi}{|\log q|})$. These special functions obey a non-trivial Yang-Baxter relation, which is proved in \cite{Jafferis:2022wez}, that ensures that the $q$-deformed Feynman rules are well-defined. Note that these $q$-deformed Feynman rules do not obey the constraint \eqref{eq:constraint}, nor do they appear to obey any other constraint.

Next, it helps to organize the calculation by introducing a fictitious parameter $\epsilon$ such that higher-point gravitational amplitudes are weighted by higher powers of $\epsilon$ (roughly speaking). We determine the couplings order by order in $\epsilon$, and at the end of the calculation we set $\epsilon \rightarrow 1$ (as well as $q \rightarrow 1$) so that the gravitational amplitudes return to their original values.

We consider two schemes for introducing the parameter $\epsilon$. One way, called the ``Selberg regulator,'' is to weight each connected $2n$-point function (given by $g^{(2 n)}_\calo$, introduced in \eqref{eq:refinedansatz}) by a factor of $\epsilon^{n-1}$. The other way, called the ``$q$-deformed regulator,'' is to weight each 6j symbol by $\epsilon$. The key point is that with either regulator, to a finite order in $\epsilon$, only finitely many of the $g^{(n)}_\calo$ functions are nonzero. Hence, we can match these connected correlators using \eqref{eq.ETHTwoMatrixModel} where only finitely many of the $G^{(n)}$ are nonzero. By working to higher orders in $\epsilon$, we can naturally construct a series representation for $G^{(n)}$. The models constructed using the Selberg and $q$-deformed regulators are respectively referred to as the Selberg and $q$-deformed matrix models. The JT limit is defined to be the $\epsilon \rightarrow 1, \, q \rightarrow 1$ limit.\footnote{In the Selberg model, the $q \rightarrow 1$ limit must precede the $\epsilon \rightarrow 1$ limit in order for the double-trumpet amplitudes to come out as desired.}

The $q$-deformed regulator is designed to match the Feynman rules of the double-scaled SYK model \cite{Cotler:2016fpe,Berkooz:2018jqr}. In the Majorana SYK model with $N$ Majorana fermions, interacting via a random  coupling involving a subset of $p$ degrees of freedom at a time, one scales $N, p \rightarrow \infty$, such that
\begin{equation}
	\lambda = \frac{2p^2}{N}\,,\qquad \qquad q=e^{-\lambda}
\end{equation}
are held finite. In addition to the Hamiltonian, one may define another operator $\calo$ that takes the same form of the Hamiltonian but with an independent set of random couplings. The number of fermions appearing in $\calo$ sets the scaling dimension $\Delta$. Correlators of $H$ and $\calo$ match those of the $q$-deformed regulator.

The Selberg regulator is designed to treat all of the gravitational Feynman diagram contributions to a single connected $n$-point function $g^{(n)}_\calo$ on an equal footing. In \cite{Jafferis:2022wez}, we present highly nontrivial results that indicate that the double-trumpet of the Selberg model agrees with \eqref{eq:dttoy}.

\subsection{Cylinder amplitudes in the matrix model}

We partially outline our techniques for computing cylinder amplitudes in the $q$-deformed and Selberg matrix models. While these models produce the same disk amplitudes in the JT limit, they return different results for the matter determinant on the double-trumpet. The model calculation we choose to present here concerns the double-trumpet two-point function, with one $\calo$ inserted into each trace. This quantity is computed in the matrix model by summing over all `t Hooft diagrams with cylinder topology and two external double-lines associated to the $\calo$ matrix. We can classify these diagrams systematically, and each class of diagrams may be computed from our knowledge of the disk correlators.

Our first class of diagrams may be summed by setting $b = a$ in \eqref{eq.MatchTwoPoint} and then integrating the remaining energy $E_a$ using the disk density of states:
\begin{equation}
	\label{eq:singleintegral}
	\int_0^\infty dE_a \,e^{S_0} \rho_0(E_a) \overline{{\cal O}_{aa} {\cal O}_{aa}},
\end{equation}
which graphically is represented as follows:
\begin{equation*}
	\includegraphics[scale=0.3]{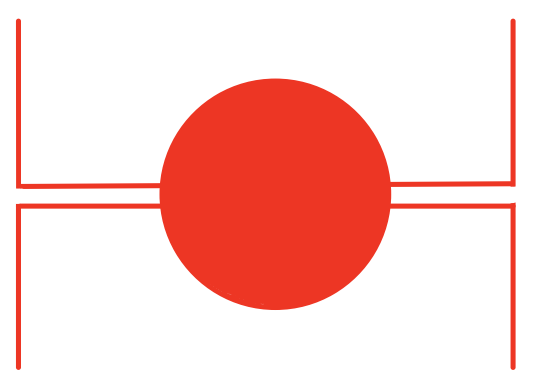} \, ,
\end{equation*}
where the top and bottom ends of the diagram are identified, so that all of the single-lines shown are connected, reflecting the fact that there is a single integral in \ref{eq:singleintegral}. It is already known \cite{Saad:2019pqd} that \eqref{eq:singleintegral} is equal to the double-trumpet two-point function without the matter determinant contribution, and we have shown that this is naturally associated to a sum over a class of `t Hooft diagrams.

The remaining `t Hooft diagrams represent non-trivial contributions of the matter determinant on the double-trumpet. For instance, let us define a blob with an ``$A$'' to represent a sum over four-point, planar, amputated `t Hooft diagrams:
\begin{equation}
		\includegraphics[scale=0.4]{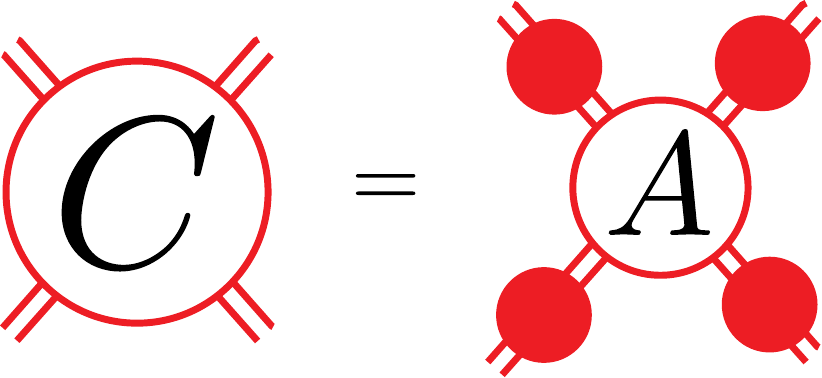} \, .
\end{equation}
We can compute another class of `t Hooft diagrams as follows:
\begin{equation}
	\label{eq:anotherclass}
		\includegraphics[scale=0.17]{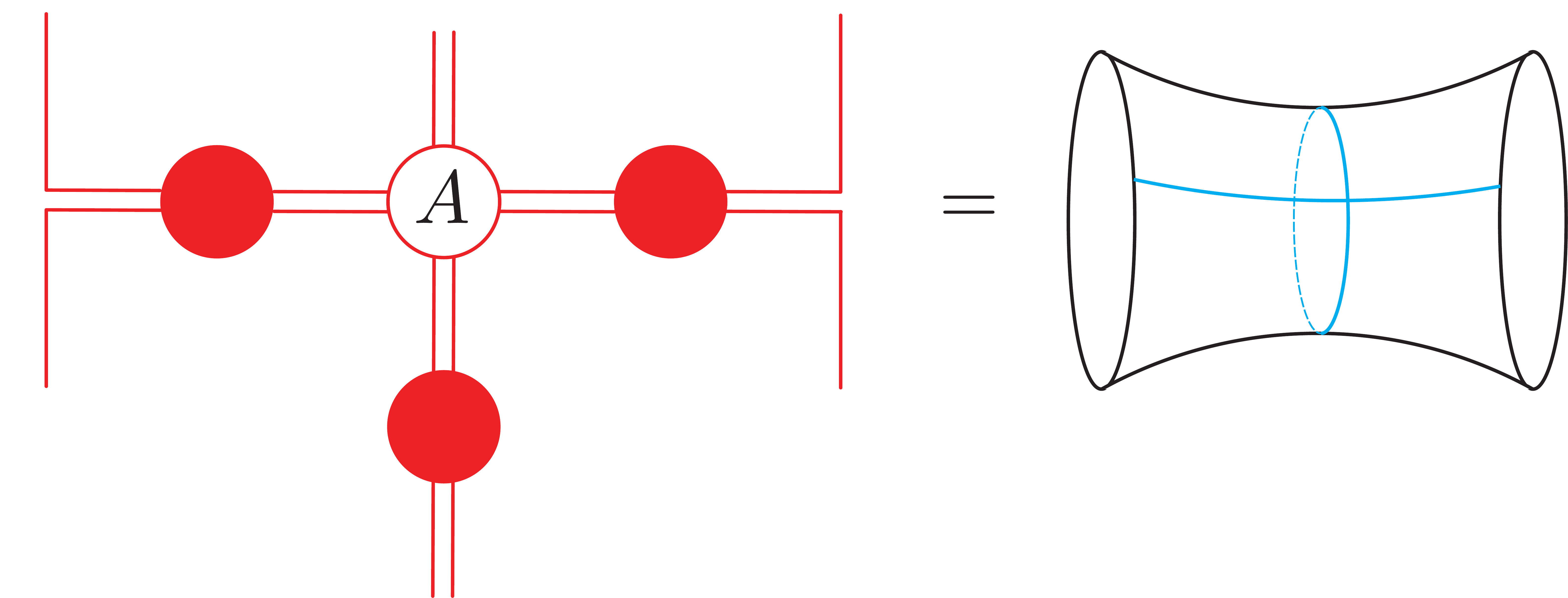} \, ,
\end{equation}
where on the left side the top and bottom of the diagram are again identified, and we have used \eqref{eq.MatchTwoPoint} and \eqref{eq.MatchFourPoint} to produce an explicit analytic expression that is graphically represented on the right side. The expression is obtainable from the gravitational Feynman rules (see the end of section \ref{sec.diskCorrelators}). Explicitly, we are computing
\begin{equation}
e^{2S_0}\int_0^\infty dE_a dE_b \, \rho_0(E_a) \rho_0(E_b) \,	e^{-3 S(\bar{E})} g^{(4)}_{\calo}(E_a,E_b,E_a,E_b) \left(~ \overline{\calo_{ab} \calo_{ba}} ~\right)^{-1}.
\end{equation}
The presence of the $\left( ~ \overline{\calo_{ab} \calo_{ba}} ~ \right)^{-1}$ reflects the fact that a red blob on the topmost vertical double-line in \eqref{eq:anotherclass} is missing, to avoid overcounting `t Hooft diagrams.

The right hand side of \eqref{eq:anotherclass} exactly computes the second term in \eqref{2pt-dt2}, and \eqref{eq:singleintegral} computes the first term in \eqref{2pt-dt2}. It is reasonable to expect that our strategy of sewing together amputated disk diagrams into cylinder diagrams will produce all of the terms in \eqref{2pt-dt2}. However, as explained in \cite{Jafferis:2022wez}, the analytic expressions for the remaining classes of `t Hooft diagrams depend on how the double-scaling limit is taken, or equivalently on whether the Selberg or $q$-deformed regulator is used.

Using the Selberg regulator, we have explicitly reproduced the third term in \eqref{2pt-dt2}, as well as the next term that corresponds to $n = 3$ in \eqref{eq:second}. We conjecture that all of the contributions from \eqref{eq:second} are matched in the Selberg matrix model. In the $q$-deformed matrix model, the results point to a different matter determinant. Instead of \eqref{eq:second}, the $q$-deformed model predicts that the matter determinant is instead
\begin{equation*}
	Z(b) = \sum_{n = 0}^\infty \left(\frac{e^{-  \Delta b}}{1-e^{-b}}\right)^n = \frac{1 - e^{-b}}{1 - e^{-b} - e^{- \Delta b}},
	\label{eq:qdefdet}
\end{equation*} 	which curiously has a Hagedorn temperature owing to the fact that the sum fails to converge for sufficiently small $b$.

Regardless of whether the matter determinant is \eqref{eq:qdefdet} or \eqref{eq:second}, the double-trumpet \eqref{eq:dttoy} is ill-defined due to the behavior of the integrand at small $b$. This implies that the gravitational theory is not UV complete, although the Selberg and $q$-deformed matrix models are operationally well-defined. As we explain in \cite{Jafferis:2022wez}, the ill-defined cylinder amplitude implies that the saddle-point that would define the genus expansion is perturbatively unstable.

\section{Discussion}
In this paper, together with its companion, \cite{Jafferis:2022wez}, we have described a way of combining the eigenstate thermalization hypothesis with certain matrix integrals into one joint framework. This allows one to interpret the ETH ansatz as arising from a joint probability distribution describing the statistics of energy levels as well as matrix elements in one unified framework. In this work we focused on two-matrix integrals because we were interested in the case of one particular operator in addition to the energy eigenvalues, but it should be clear that adding further operators is possible and will lead to multi-matrix integrals. The structure of these multi-matrix integrals is related to free probability theory, which was invoked in the ETH context previously by \cite{Pappalardi:2022aaz}. It would be interesting to investigate this connection further.

We have further expanded on what is presumably the simplest instance of our ETH matrix model, namely the case of thermal mean field theory. It is striking that even in this simplest context the matrix model is strongly non-Gaussian, even though as seen in the energy eigenbasis all non-Gaussian contributions are entropically suppressed.\footnote{In \cite{Hunter-Jones:2017raw}, the Gaussian ETH was numerically verified in the SYK model. It would be interesting if one could measure the entropically suppressed non-Gaussianities in a more sensitive study.} This is of course a necessary feature of  ETH matrix models in general, and compatible with the statistical physics approach of \cite{Foini:2018sdb}.

More generally, quantum chaotic systems are not expected to be described exactly by a matrix integral, rather they approach such behavior at late times. Usually the timescale at which a matrix theory description accurately captures a quantum chaotic Hamiltonian is referred to as the Thouless or ergodic time scale.  In fact, a more precise definition of this time scale also demands that the statistics of both matrix elements as well as eigenvalues approach those of Gaussian random matrices (see \cite{Wang:2021mtp} for an in-depth discussion). In this work we formulated matrix models which apply at (much) earlier time scales precisely by incorporating non-Gaussian statistics into the joint probability distribution of matrix elements and eigenvalues. We extend the applicability of the ETH matrix model beyond the Thouless time by adding more information about the physical system in the form of non-Gaussian terms in the joint potential, a perspective that is made very clear in the constraint matrix model approach we outlined above. In order to generalize this idea, one should introduce a {\it matrixiziation timescale} $t_M$, beyond which a system is well described by an ETH matrix model. By introducing further constraints along the lines of \eqref{eq:constraint}, one obtains ETH models whose $t_M$ is pushed further and further towards early time. This procedure is very much in the spirit of an {\it effective matrix model} (see \cite{Jafferis:2022wez}), whose region of validity can be extended by adding more and more UV information, paralleling the procedure one would follow in effective field theory.  It is intriguing that certain systems, such as pure JT gravity or JT gravity with matter as well as the double-scaled SYK model exhibit a matrixization timescale $t_M$ that formally tends to zero.  It would be interesting to ask under what conditions this can happen more generally, and furthermore whether there may be lower bounds on $t_M$ in higher dimensional (holographic) theories.

For a more extensive discussion of open questions and for future directions related to our ETH matrix model, the reader is referred to the companion paper \cite{Jafferis:2022wez}.

\paragraph{Acknowledgements}

We would like to thank Nick Agia, Alex Belin, Noam Chai, Jan de Boer, Anatoly Dymarsky, Lorenz Eberhardt, Akash Goel, Tom Hartman, Clifford Johnson, Zohar Komargodski, Henry Lin, Juan Maldacena, Dalimil Maz\'{a}\v{c}, Vladimir Narovlansky, Pranjal Nayak, Joaquin Turiaci, and Herman Verlinde for stimulating discussions. This work was performed in part at Aspen Center for Physics, which is supported by National Science Foundation grant PHY-1607611. This work has been partially supported by the DOE through the grant DE-SC0007870, SNF through Project Grants 200020 182513, as well as the NCCR 51NF40-141869 The Mathematics of Physics (SwissMAP). The work of BM was supported by a grant from the Simons Foundation (651444, BM) and NSF grant PHY-2014071.

\bibliographystyle{JHEP}
\nocite{}
\bibliography{thebibliography}
\end{document}